\documentclass[aip,jcp]{revtex4-1}

\usepackage{epsfig}
\usepackage{latexsym}
\usepackage{amsmath}
\usepackage{amsfonts}
\usepackage{amssymb}
\usepackage{amsbsy}
\usepackage{subfigure}
\usepackage{bm}
\usepackage{color}
\usepackage{dcolumn}
\usepackage{hyperref}

\graphicspath{{./figures/}}

\newcommand{\lla}{\left\langle}
\newcommand{\rra}{\right\rangle}

\newcommand{\red}{\color{red}}
\newcommand{\blue}{\color{blue}}
\newcommand{\green}{\color{green}}

\begin{document}
\title{Semiflexible polymers under external fields
confined to two dimensions}
\author{A. Lamura}
\email[]{a.lamura@ba.iac.cnr.it}
\affiliation{
Istituto Applicazioni Calcolo, CNR,
Via Amendola 122/D, 70126 Bari, Italy}
\author{R. G. Winkler}
\email[]{r.winkler@fz-juelich.de} \affiliation{Theoretical Soft
Matter and Biophysics, Institute for Advanced Simulation,
Forschungszentrum J\"{u}lich, 52428 J\"{u}lich, Germany}
\date{\today}
\begin{abstract}
The non-equilibrium structural and dynamical  properties
of semiflexible polymers confined to two dimensions
are investigated
by molecular dynamics simulations. Three different scenarios are
considered: The force-extension relation of tethered polymers, the
relaxation of an initially stretched semiflexible polymer, and
semiflexible polymers under shear flow. We find quantitative
agreement with theoretical predictions for the force-extension
relation and the time dependence of the entropically contracting
polymer. The semiflexible polymers under shear flow exhibit
significant conformational changes at large shear rates, where
less stiff polymers are extended by the flow, whereas rather stiff
polymers are contracted. In addition, the polymers are aligned by
the flow, thereby the two-dimensional semiflexible polymers behave
similarly to flexible polymers in three dimensions. The tumbling
times display a power-law dependence at high shear rate rates
with an exponent comparable to the one of flexible polymers in 
three-dimensional systems.
\end{abstract}
\maketitle

\section{Introduction}

Semiflexibility is a characteristic property of a broad range of
biological polymers. Prominent examples are DNA, filamentous
actin, microtubules, or viruses such as fd-viruses
\cite{janm:95,bust:94,mark:95,wilh:96,ripo:08}. The rigidity is
fundamental for their biological functions. For example, the DNA
persistence length strongly affects its packing in the genome or
inside a virus capsid. Actin filaments are an integral part of
the cytoskeleton and their rigidity determines its particular
mechanical properties. Hence, considerable effort has been devoted
to unravel the mechanical and dynamical properties of semiflexible
polymers
\cite{maed:84,arag:85,farg:93,goet:96,harn:96,wilh:96,ever:99,samu:02,lego:02,wink:03,carr:04,petr:06,wink:06,wink:07.1}.

Advances in single-molecule spectroscopy prompted experimental and
theoretical studies of non-equilibrium properties of semiflexible
polymers
\cite{perk:95,perk:97,quak:97,wink:99,ledu:99,smit:99,schr:05,gera:06}.
Fluorescence microscopy studies on single DNA molecules in shear
flow reveal large conformational changes and an intriguing
dynamics, denoted as tumbling motion
\cite{ledu:99,smit:99,schr:05,gera:06}. This implies specific
non-equilibrium conformational, dynamical, and rheological
properties, which have been analyzed experimentally
\cite{smit:99,schr:05,teix:05,schr:05_1,doyl:00,lado:00,gera:06},
theoretically
\cite{john:87,brun:93,carl:94,wang:01,woo:03,dubb:06,bird:87,oett:96,dua:00,rabi:89,prak:01,wang:89,bald:90,gana:99,subb:95,prab:06,wink:06_1,munk:06,wink:10},
and by computer simulations
\cite{liu:89,cela:05,hur:00,jose:08,he:09,knud:96,lyul:99,jend:02,hsie:04,liu:04,pami:05,schr:05,send:08,zhan:09,pier:95,aust:99,grat:05,ryde:06,ripo:06,koba:10,huan:10,huan:11,huan:12}.

These studies typically consider semiflexible polymers in three-dimensional
space. Much less attention has been payed to polymers
in two dimensions, although we may expect to see particular
features in their equilibrium and non-equilibrium dynamical
properties. Two-dimensional behavior is realized for strongly
adsorbed polymers at, e.g., a solid surface, a membrane, or at the
interface between immiscible fluids. Experiments reveal a strong
dependence of the diffusive dynamics of adsorbed polymers on the
underlaying substrate \cite{maie:99,sukh:00}. Moreover,
theoretical and simulation studies predict a strongly correlated
dynamics in two-dimensional polymer melts \cite{witt:10}. Little
is known about the non-equilibrium properties of polymers in two
dimensions. Here, we refer to  recent simulation studies of
end-tethered semiflexible polymers, where the central monomer is
periodically excited \cite{chat:07,zhan:11}. These simulations
find a crossover from a limit cycle to an aperiodic dynamics with
increasing stiffness.

There are two major differences to three-dimensional systems. First
of all, excluded volume interactions play a more pronounced role.
The non-crossability leads, e.g., to a segregation of polymers in
two dimensions \cite{witt:10}. We expect a strong impact of these
interactions on non-equilibrium properties too.  Secondly,
hydrodynamic interactions can be neglected under certain
circumstances \cite{maie:02}. This applies to strongly adsorbed
polymers, where the polymer-substrate interaction dominates the
dynamics of the polymer. It is certainly not appropriate for
polymers confined at fluid-fluid interfaces.

In this article we investigate the non-equilibrium structural and
dynamical properties of  semiflexible polymers  by computer
simulations. As discussed above, we assume that the local polymer
friction is determined by its interaction with the substrate and,
hence, neglect hydrodynamics. Thus, we exploit the
Brownian multiparticle collision (B-MPC) dynamics approach described in
Refs.~\onlinecite{ripo:07,gomp:09,wink:12}. By varying the chain
stiffness, we gain insight into the dependence of the polymer
properties on stiffness. Moreover, by comparison with existent
results on three-dimensional systems, we uncover specific effects
of the reduced dimensionality.

Three different situations are considered. We briefly touch the
force-extension relation of a semiflexible polymer and show that
it is well described by theory \cite{mark:95,wink:03}. In
addition, we examine the end-to-end vector relaxation behavior of
initially stretched polymers. We find excellent agreement with the
power-law dependence obtained in experiments \cite{maie:02}. This
suggests that our model is a useful coarse-grained representation
of a DNA molecule, at least for the considered properties. The
major focus of the paper is on the non-equilibrium properties of
semiflexible polymers under shear flow. We discuss a broad range
of structural and dynamical quantities and stress the universal
character and/or their particular, two-dimensional features.

The paper is organized as follows. In Sec.~\ref{sec:model} the
polymer model is described and the simulation method is
introduced. In Sec.~\ref{sec:polymer_ext} the force-extension
relation of a tethered polymer in a uniform external field is
discussed. Section~\ref{sec:relax} presents results on the time
dependent relaxation behavior of a stretched semiflexible polymer.
The structural properties of free polymers under shear flow are
discussed in Sec.~\ref{sec:shear}, and their tumbling dynamics is
analyzed in Sec.~\ref{sec:tumbling}. Finally,
Sec.~\ref{sec:conclusion} summarizes our findings.

\section{Model and Method} \label{sec:model}

The polymer is modeled as a linear chain composed of $N$ beads of
mass $M$. Its intramolecular interactions are described by the
potential $U=U_{bond}+U_{bend}+U_{ex}$. Successive beads are
linked by the harmonic bond potential
\begin{equation} \label{bond}
U_{bond}=\frac{\kappa_h}{2} \sum_{i=1}^{N-1}
(|{\bm r}_{i+1}-{\bm r}_{i}|-r_0)^2 ,
\end{equation}
where ${\bm r}_i$ is the position vector of bead  $i$
($i=1,\ldots,N$), $\kappa_h$ is the spring constant, and $r_0$ the
bond length. The bond bending potential
\begin{equation}
U_{bend}=\kappa \sum_{i=1}^{N-2} (1-\cos \varphi_{i})
\label{bend}
\end{equation}
accounts for the bending stiffness of the polymer, with $\kappa$
the bending rigidity and $\varphi_{i}$ the angle between two
consecutive bond vectors. In the semiflexible limit $\kappa \to
\infty$,  the bending stiffness is related to the persistence
length by $L_p=2 \kappa r_0/ k_B T$, where $k_B T$ is the thermal
energy, with $T$ the temperature and $k_B$ Boltzmann's constant.
Excluded-volume interactions are ensured by the shifted and
truncated Lennard-Jones potential
\begin{equation}
U_{ex} =
4 \epsilon \Big [ \Big(\frac{\sigma}{r}\Big)^{12}
-\Big(\frac{\sigma}{r}\Big)^{6} +\frac{1}{4}\Big] \Theta(2^{1/6}\sigma -r) ,
\label{rep_pot}
\end{equation}
where $r$ denotes the distance between two non-bonded beads and
$\Theta(r)$ is the Heaviside function ($\Theta(r)=0$ for $r<0$ and
$\Theta(r)=1$ for $r\ge 0$). The dynamics of the beads is
described by Newton's equations of motion, which we integrated by
the velocity-Verlet algorithm with time step $\Delta t_p$
\cite{swop:82,alle:87}.

The polymer is coupled to a Brownian heat bath, which we implement
via the B-MPC approach
\cite{ripo:07,gomp:09,kiku:03}. Hence, no hydrodynamic
interactions are taken into account. In B-MPC, a bead performs
stochastic collisions with a phantom particle which  mimics a
fluid element of a certain size. The momentum of the phantom
particle is taken from a Maxwell-Boltzmann distribution of
variance $M k_B T$ and mean given by the average momentum of the
fluid field, which is zero at rest or $(M\dot{\gamma}
y_{i},0)^T$ in the case of an imposed shear flow of shear rate
$\dot \gamma$ in the $xy-$plane. For the stochastic process
itself, we apply the stochastic rotation dynamics
realization of the MPC method \cite{ihle:01,lamu:01,gomp:09}. Here, the
relative velocity of a polymer bead, with respect to the
center-of-mass velocity of the bead and the associated phantom
particle,  is rotated in the $xy-$plane by angles $\pm \alpha$. The
time interval between collisional interactions is $\Delta t$,
which is larger than the time step $\Delta t_p$.

The simulations are performed for the parameters $\alpha=130^{o}$,
$\Delta t=0.1 t_u$, where the time unit is $t_u=\sqrt{m r_0^2/(k_B
T)}$, $M =5m$, $\kappa_h r_0^2/(k_B T)=4 \times 10^3$, $\epsilon /
(k_B T)=1$, $\sigma=r_0$, $N=51$ so that polymer length is $L=50
r_0$, and $\Delta t_p=10^{-2} \Delta t$. With this choice for
$\kappa_h$, the length of the polymer is kept constant within
$1\%$ for all systems.

\section{Polymer Force-Extension Relation in Uniform Field} \label{sec:polymer_ext}

We consider a single tethered polymer with its endpoint ${\bm
r}_1$ fixed at ${\bm r}_1 =0$ without additional restrictions on
the orientation of the first bond. Every monomer is subjected to
the external force $F$ along the $x$-direction of the Cartesian
reference frame, e.g., due to an external electric field $E_x$,
where the force is $F=qE_x$, with $q$ the electric charge of the
monomer. As is well known, the polymer is stretched along the
force direction, where the extension increases non-linearly with
increasing force \cite{lamu:01.1,hori:07}. Theoretical
calculations based on the Kratky-Porod \cite{krat:49} wormlike
chain model predict the asymptotic dependence
\begin{align} \label{force_extens}
\frac{x_{N}}{L} = 1-\left( \frac{k_B T}{2 L_p F (N-1)}
\right)^{1/2}
\end{align}
for $|x_{N}| \to L$.

\begin{figure}
\begin{center}
\includegraphics*[width=.45\textwidth,angle=0]{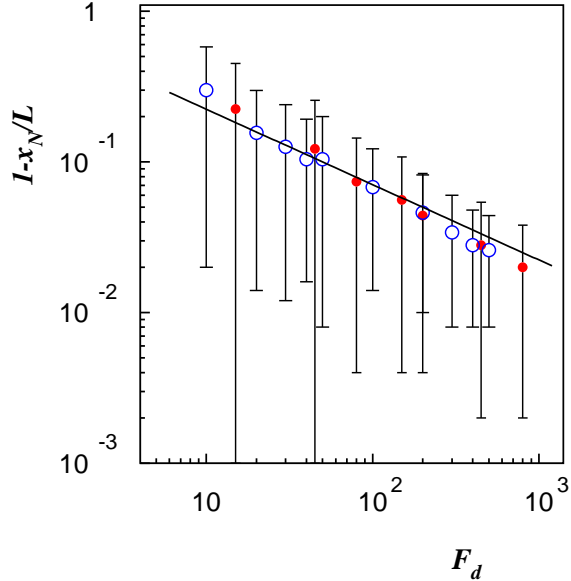}
\caption{Polymer extension along the direction of the external force
for the persistence lengths  $L_p/L= 2 ({\blue \circ}), \ 4 ({\red \bullet})$
as a function
of the dimensionless force $F_d=L_p F (N-1)/(k_B T)$.
The full line represents  Eq.~(\ref{force_extens}) and has the
slope $-1/2$.
 \label{fig:extens}}
\end{center}

\end{figure}

Simulation results are presented in Fig.~\ref{fig:extens} for the
persistence lengths  $L_p/L= 2, \ 4$. They agree very well with
the theoretical prediction  (\ref{force_extens}). This confirms
that the model represents a continuous semiflexible polymer over
the presented range of forces. Due to the discrete nature of the
model, deviations will appear from the predictions of  continuous
semiflexible polymers \cite{mark:95,odij:95,wink:03,hori:07} for
large forces, as shown in Refs.~\onlinecite{lamu:01.1,liva:03,rosa:03}, 
and a crossover will occur from the force-extension relation of a
semiflexible model to that of a freely jointed chain.
The effects of attractive interactions between non consecutive
nearest neighbor beads \cite{rosa:03}
on the described picture might be interesting to be investigated in the
future.

\section{Relaxation of Stretched Polymer} \label{sec:relax}

\begin{figure}
\begin{center}
\includegraphics*[width=.45\textwidth,angle=0]{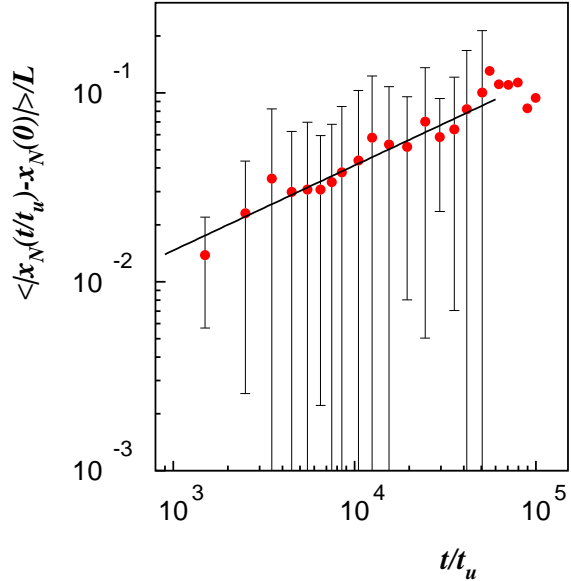}
\caption{Relaxation of the end-to-end distance along the stretching
direction  $|x_{N}(t/t_u)-x_{N}(0)|$ of a semiflexible polymer with $L_p/L=2$.
The line indicates a fit in the time range $t/t_u=10^3 - 50 \times 10^3$
and has the slope $0.45$. \label{fig:relax}
}
\end{center}
\end{figure}

Releasing the force on a stretched polymer leads to its collapse
into an equilibrium conformational state. This relaxation exhibits
a characteristic time dependence. Figure~\ref{fig:relax} displays
the relaxation behavior of the $x$-component of the end-to-end
vector of a stretched semiflexible polymer. The initial average
stretching along the $x-$direction is $x_{N}(0)/L = 0.945$, induced by the force $FL_p(N-1)/(k_B
T) = 200$. We observe a power-law decrease of the extension
according to $|x_{N}(t)-x_{N}(0)|/L \sim t^{\gamma}$ over a broad
time scale.  A fit of the data, which are averages over 20
independent realizations, yields the exponent $\gamma = 0.45 \pm
0.03$. This value is in remarkable agreement with the exponent
$0.46 \pm 0.08$ found in two-dimensional experiments
\cite{maie:02}.

\section{Semiflexible Polymer under Shear Flow} \label{sec:shear}

In our study of semiflexible polymers under shear flow in
two-dimensional space, we consider the persistence lengths
$L_p/L=0.1, 0.4, 2, 10$. The corresponding equilibrium end-to-end
vector relaxation times are $\tau_0 / t_u \simeq (161, \ 370, \
676,$ and $707$) $\times 10^3$, respectively. The
strength of the flow is characterized by the Weissenberg number $Wi=\dot{\gamma}
\tau_0$ in the range $1 \le Wi \le 800$.

\subsection{Conformational Properties}

\subsubsection{End-to-End Vector}

\begin{figure}
\begin{center}
\includegraphics*[width=.6\textwidth,angle=0]{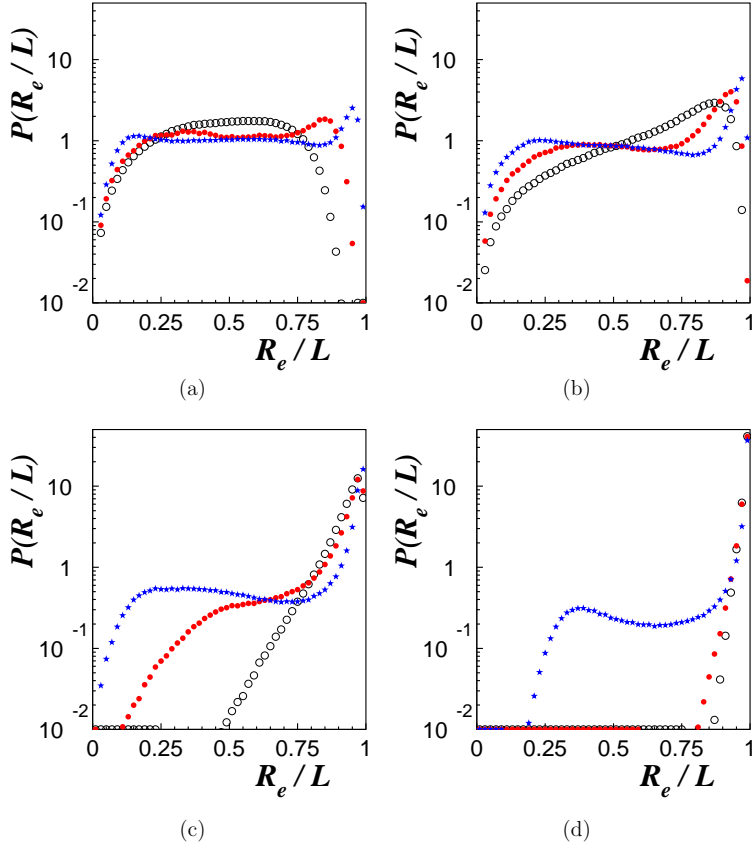}
\caption{ Probability distributions of the polymer end-to-end
distance $R_e =|{\bm r_{N}} - {\bm r}_1|$ for the  Weissenberg numbers $Wi=
8 (\circ), \  80 ({\red \bullet}), $ and $800 ({\blue \star})$ and
$L_p/L=0.1 (a), \ 0.4 (b), \ 2 (c), \ 10 (d)$.
\label{pdflen}}
\end{center}
\end{figure}

Probability distribution functions (PDFs) of the polymer
end-to-end distance $R_e =|{\bm r_{N}} - {\bm r}_1|$ are presented
in Fig.~\ref{pdflen} for the Weissenberg numbers $Wi=8, \ 80, \
800$ and the various persistence lengths. As shown in
Fig.~\ref{pdflen}(a), for small persistence lengths, the polymers
are able to assume coil-like conformations, which give nearly
constant PDFs over a wide range of end-to-end distances. Very
small distances are suppressed by excluded-volume interactions and
large distances are rarely sampled due to entropic penalties.
However, larger shear rates lead to a sampling of large $R_e$
values.  At low shear rates, an increasing persistence length
naturally leads to a preference of large $R_e$ values. Shear,
however, leads to an opposite behavior. At a given $L_p/L$, an
increasing shear rate gives rise to an increase in the probability
distribution at smaller end-to-end distances. The effect becomes
more pronounced for larger stiffnesses (cf.
Fig.~\ref{pdflen} (c) and (d)). This is in agreement with the predictions
of Ref.~\onlinecite{wink:10} that semiflexible polymers under shear flow
behave more and more like flexible polymers with increasing
Weissenberg number.

\begin{figure}[ht]
\begin{center}
\includegraphics*[width=.45\textwidth,angle=0]{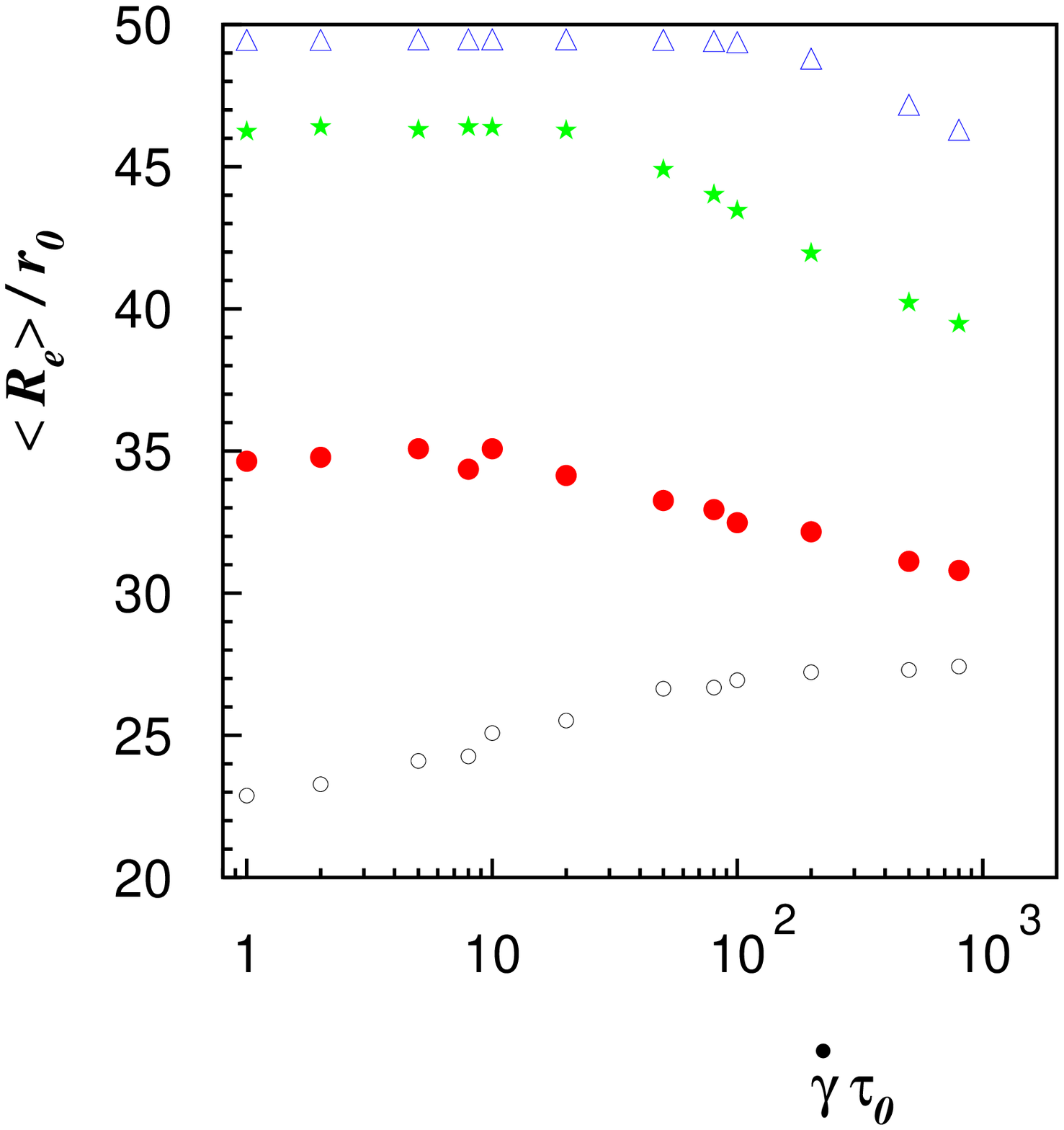}
\caption{Mean values $\lla R_e \rra$ of the end-to-end distance as
function of the Weissenberg number for the persistence lengths
$L_p/L=0.1 (\circ), \ 0.4 ({\red \bullet}), \ 2 ({\green \star}),$ and
$10 ({\blue \triangle})$.
\label{rmed}
}
\end{center}
\end{figure}

Figure~\ref{rmed} displays the mean end-to-end distances as
function of $Wi$. For every shear rate, $\lla R_e \rra$ is smaller
for the more flexible polymer. However, $\lla R_e \rra$ increases
for flexible polymers, whereas it decreases for the stiffer ones.
As predicted by theory \cite{wink:10}, we expect that the
end-to-end distances become similar for all stiffnesses in the
asymptotic limit $Wi \to \infty$. As the figure clearly reveals,
in the stationary non-equilibrium state, a semiflexible polymer is
never fully stretched. This has also been observed in simulations
of flexible polymers \cite{huan:10,huan:11} and in experiments on
DNA molecules \cite{schr:05,schr:05_1}.

\begin{figure}[t]
\begin{center}
\includegraphics*[width=.45\textwidth,angle=0]{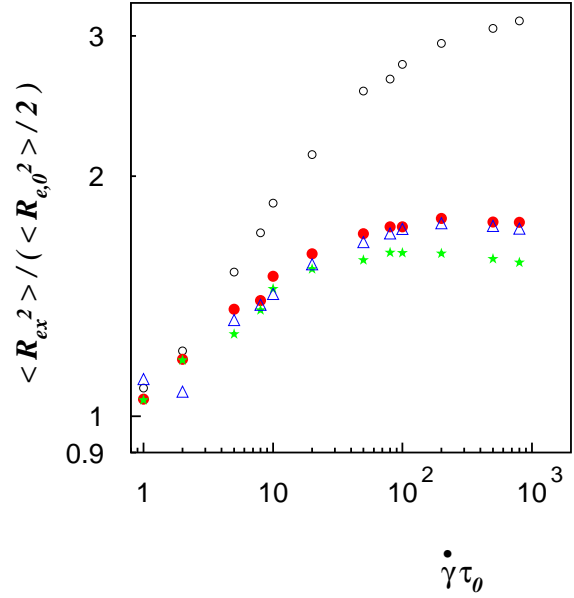}
\caption{Mean square end-to-end distances along the
flow direction as function of the
Weissenberg number for the persistence lengths
$L_p/L=0.1 (\circ), \ 0.4 ({\red \bullet}), \ 2 ({\green \star}),$ and
$10 ({\blue \triangle})$.
$\lla {\bm R}_{e,0}^2\rra$ is the mean square end-to-end distance at
equilibrium.
\label{lenx_bis}}
\end{center}
\end{figure}

Mean square end-to-end distances $\lla R^2_{ex} \rra$ along the
flow direction are displayed in Fig.~\ref{lenx_bis}.  They also
show that semiflexible polymers are never fully stretched.
Moreover, the various curves reveal a very weak persistence length
dependence for polymers with $L_p/L \gtrsim 0.4$. They closely
follow the same Weissenberg number dependence. This has been
predicted in Ref.~\onlinecite{wink:10} and is related to the fact that
the rather stiff polymers become first aligned with the flow at
moderate shear rates. Only at larger shear rates, deformation sets
in. This is also evident from Fig.~\ref{rmed}, which clearly
exhibits a dependence of the ``critical'' Weissenberg number on
the persistence length, above which $\lla R_e \rra$ decreases with
increasing shear rate. Below the critical value, the polymers are
aligned by the flow and above, in addition, they are deformed
\cite{wink:10}.

\begin{figure}[t]
\begin{center}
\includegraphics*[width=.45\textwidth,angle=0]{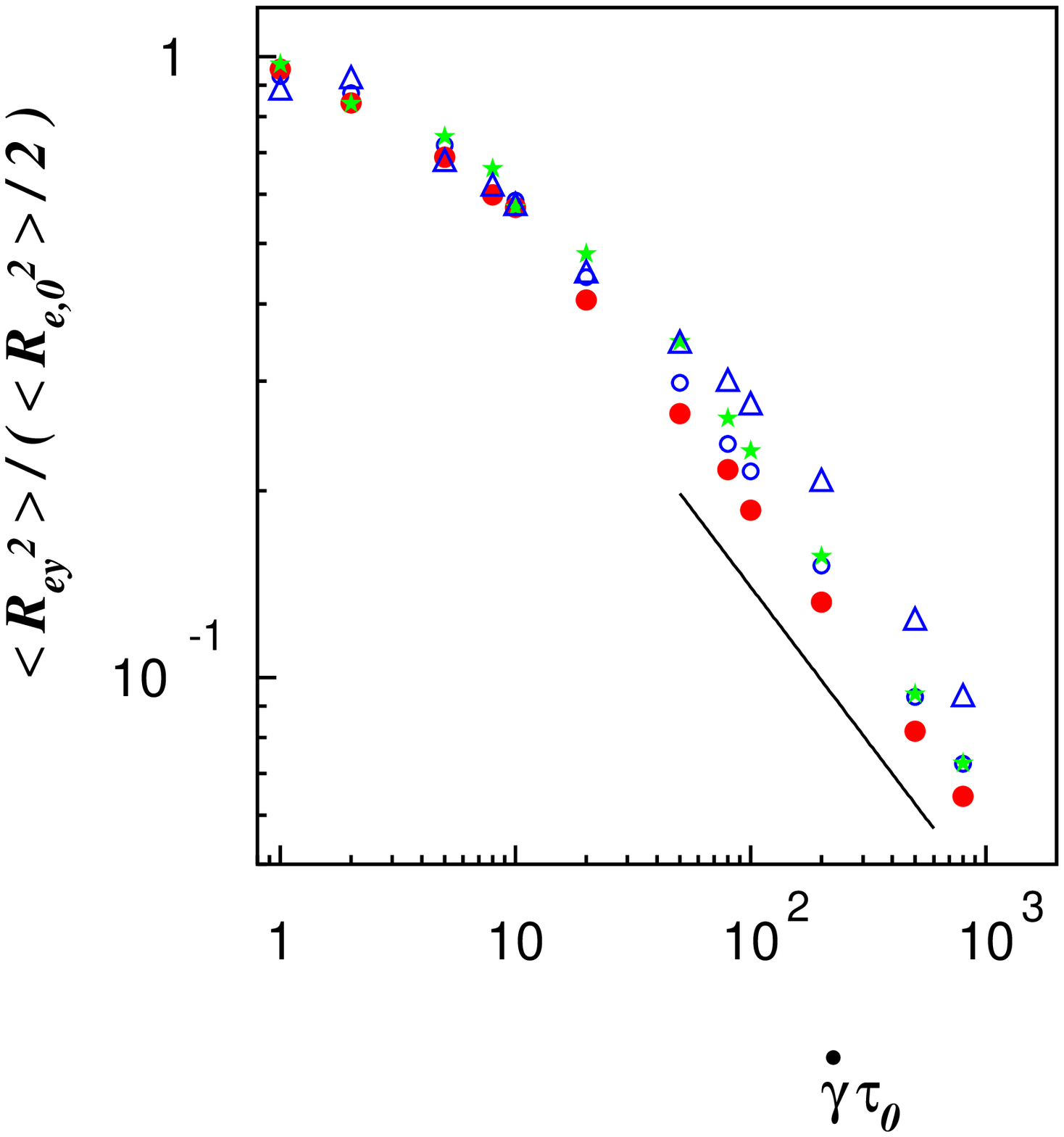}
\caption{Mean square end-to-end distances along the
gradient direction as function of the
Weissenberg number for the persistence lengths
$L_p/L=0.1 (\circ), \ 0.4 ({\red \bullet}), \ 2 ({\green \star}),$ and
$10 ({\blue \triangle})$.
The full line has the slope $-1/2$.
$\lla {\bm R}_{e,0}^2\rra$ is the mean square end-to-end distance at
equilibrium.
\label{leny_bis}}
\end{center}
\end{figure}

In Fig.~\ref{leny_bis} mean square end-to-end distances $\lla
R^2_{ey} \rra$ are shown along the gradient direction. The
polymers of the various stiffnesses shrink transverse to the flow
direction. Thereby, we observe a slight dependence on persistence
length over the considered range. The decay at larger $Wi$ can
approximately be described by the power-law $\lla R_{ey}^2 \rra
\sim Wi^{-\nu}$, with $\nu \approx 1/2$. This dependence is
consistent with the decay of the radius-of-gyration tensor of
three-dimensional systems \cite{huan:10,schr:05_1,hur:00}.
However, theoretically an exponent $\nu =2/3$ is expected
\cite{wink:06,wink:10}, which seems to be reached for much higher
Weissenberg numbers in Ref.~\onlinecite{schr:05}. Hence, the exponent
$\nu=1/2$ could characterize a crossover behavior only.

\subsubsection{Bond Angle}

\begin{figure}
\begin{center}
\includegraphics*[width=.6\textwidth,angle=0]{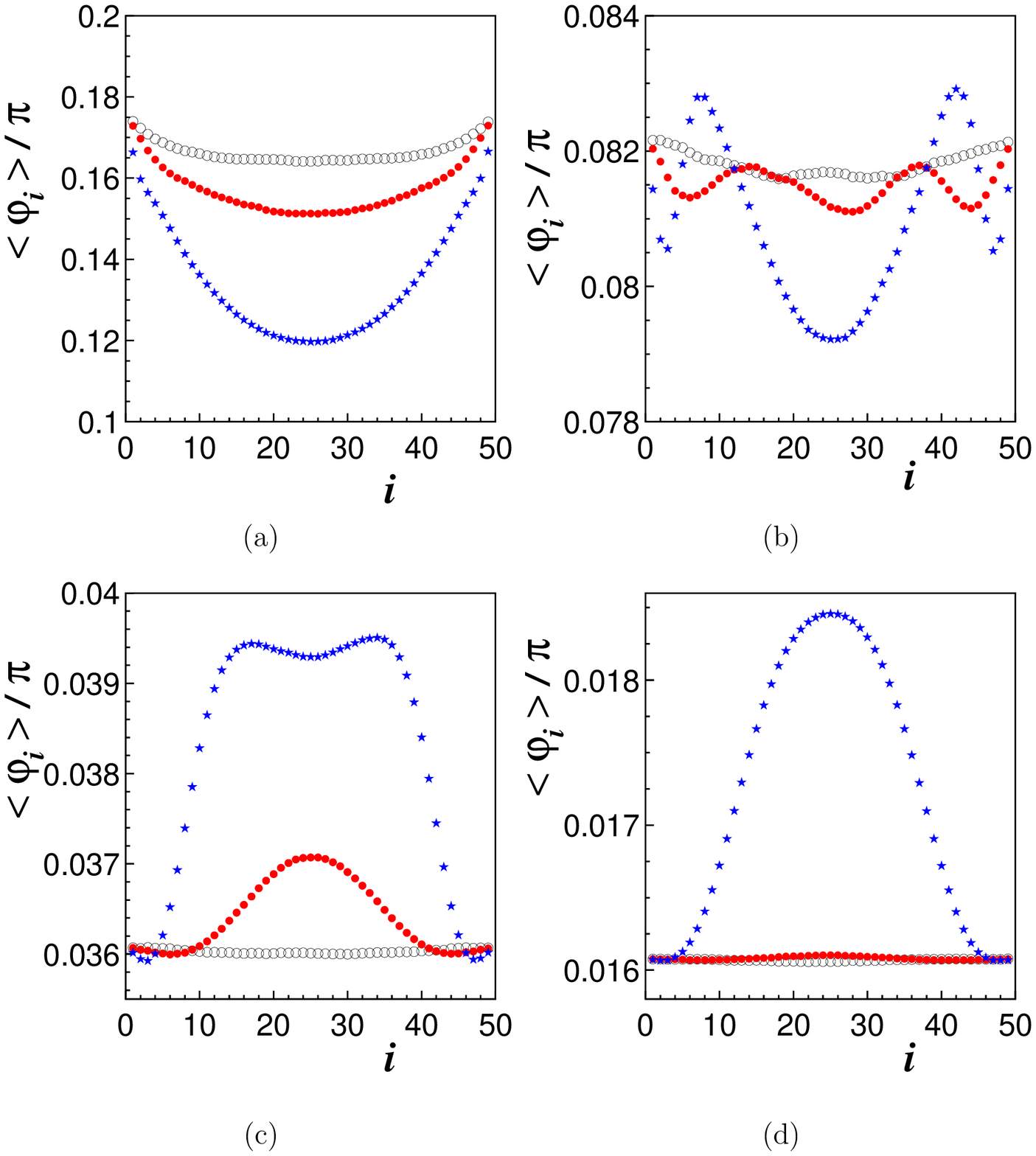}
\caption{Average local bond angle $\lla \varphi_i \rra$ along the polymer
contour for the
Weissenberg numbers $Wi =8 (\circ), \ 80, \
({\red \bullet}),$ and
$800 ({\blue \star})$ and the persistence lengths $L_p/L=0.1 (a), \
0.4 (b), \ 2 (c),$ and $10 (d)$.
\label{alphai}}
\end{center}
\end{figure}

To further characterize the polymer conformational properties, we
present in Fig.~\ref{alphai} the average bond angles $\lla \varphi_i
\rra$ (\ref{bend}) between successive bond vectors along the
semiflexible polymers. As expected, the angles $\lla \varphi_i
\rra$ decrease with increasing persistence length and are close to
zero for $L_p/L=10$. At small persistence lengths, the values of
$\lla \varphi_i \rra$ decrease with increasing shear rate,
specifically toward the middle of the chain, due to polymer
stretching by the flow. The situation is reverted at large
persistence lengths, where the angles $\lla \varphi_i \rra$ increase with
the shear rate.

\begin{figure}
\begin{center}
\includegraphics*[width=.45\textwidth,angle=0]{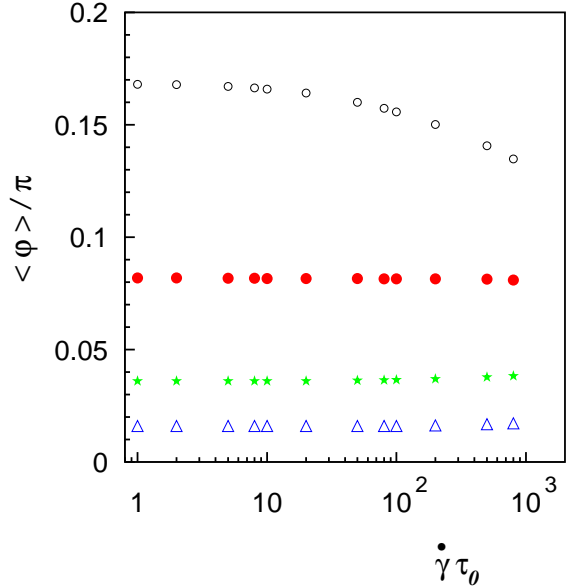}
\caption{Average bond angles $\lla \varphi \rra$ (\ref{avr_bond_ang})
as function of the
Weissenberg number for the persistence lengths
$L_p/L=0.1 (\circ), \ 0.4 ({\red \bullet}), \ 2 ({\green \star}),$ and
$10 ({\blue \triangle})$.
\label{alpha}}
\end{center}
\end{figure}

Average bond angles
\begin{align} \label{avr_bond_ang}
\lla \varphi \rra = \frac{1}{N-2} \sum_{i=1}^{N-2} \lla \varphi_i \rra
\end{align}
are displayed in Fig.~\ref{alpha} as function of Weissenberg
number and for various persistence lengths. Evidently, the mean
values are independent of shear rate for the larger persistence
lengths. Only for the considered most flexible polymer a decrease
of $\lla \varphi \rra$ is found as already expected from
Fig.~\ref{alphai}.

This minor change in the bond-bond orientational behavior is
surprising in the light of the decreasing mean end-to-end distance
(cf. Fig.~\ref{rmed}). This is explained on the one hand by the
nearly rigid rod-like rotation of the semiflexible polymers at
lower Weissenberg numbers and on the other hand by the formation
of U-shaped conformations with only small and local bending of the
polymer, as reflected in Fig.~\ref{alphai}(c) and (d) at higher
values of $Wi$. Typical conformations of stiff polymers at low and
high Weissenberg numbers are presented in Fig.~\ref{confpolym}
(see also the movies in the supplementary material \cite{supp}).

\begin{figure}
\begin{center}
\includegraphics*[width=.6\textwidth,angle=0]{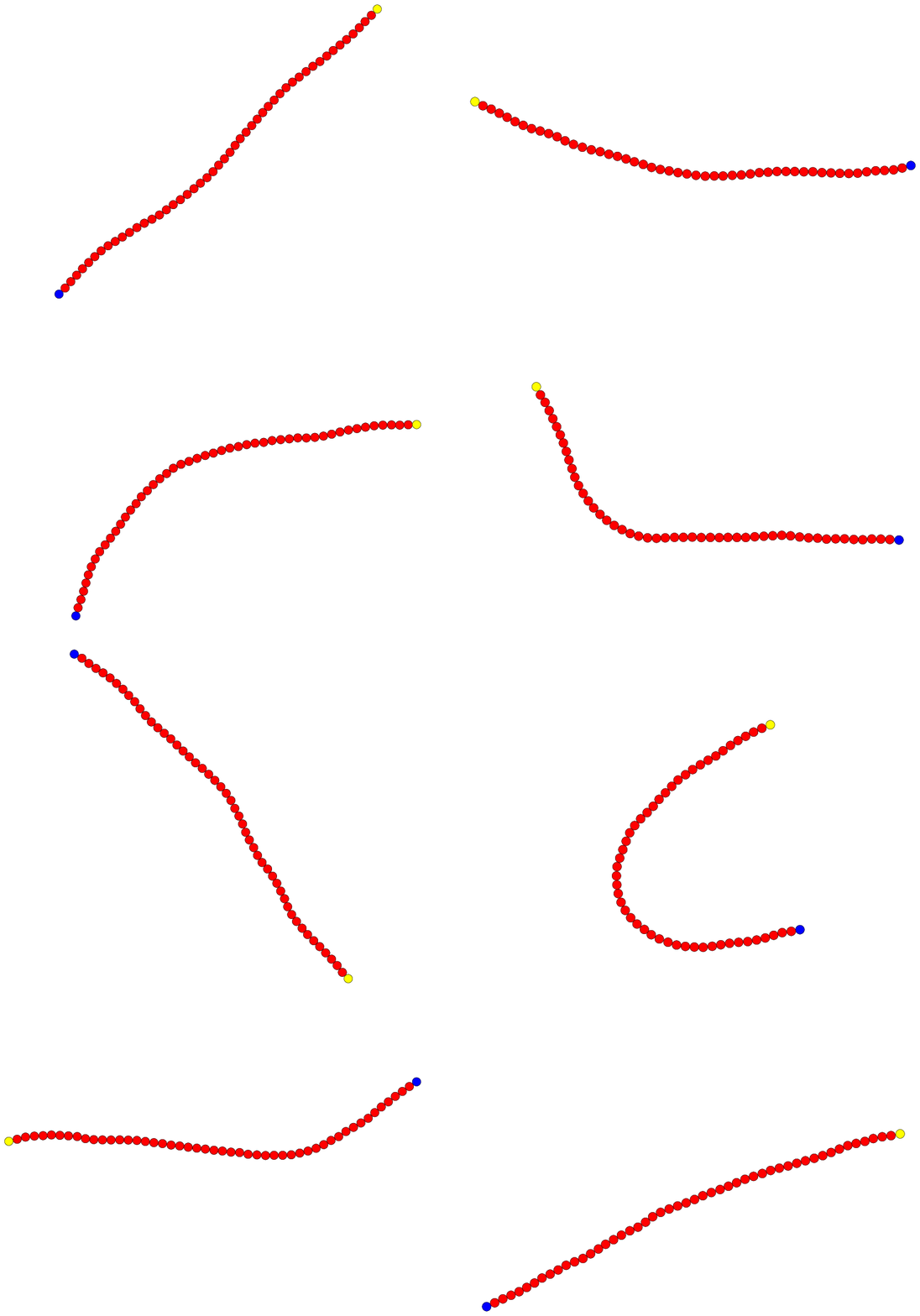}
\caption{Typical conformations at consecutive times (from top to bottom)
of a semiflexible polymer ($L_p/L=10$) at $Wi=8$ (left panel) and $Wi=800$ (right panel).
\label{confpolym}}
\end{center}
\end{figure}

\subsection{Alignment}

\begin{figure}
\begin{center}
\includegraphics*[width=.6\textwidth,angle=0]{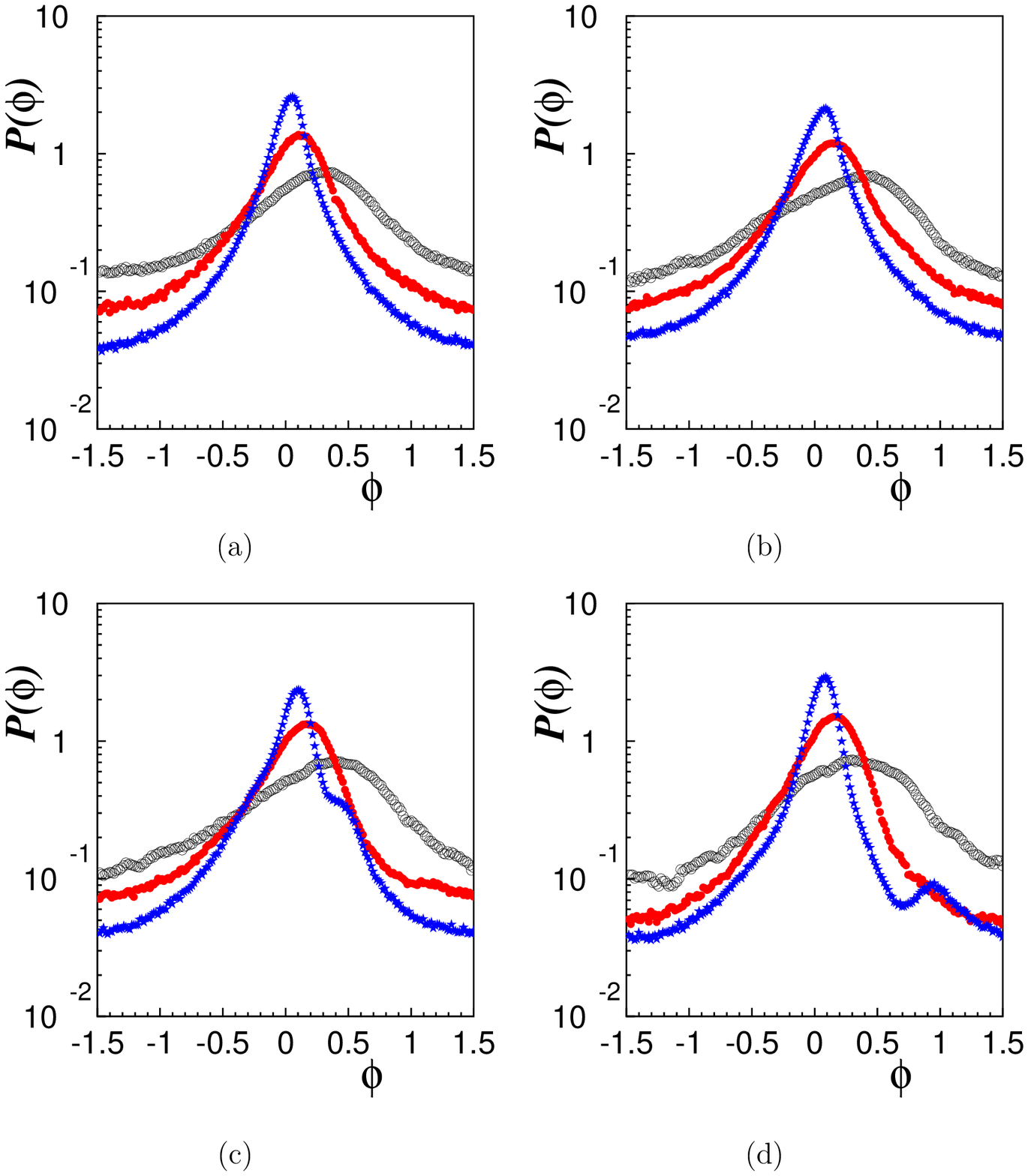}
\caption{Probability distributions of the angle $\phi$
for the Weissenberg numbers $Wi
=8 (\circ), \ 80 ({\red \bullet}),$ and $ 800 ({\blue \star})$ and the
persistence lengths $L_p/L=0.1 (a), \ 0.4 (b), \ 2 (c),$ and $10 (d)$.
\label{pdfang}}
\end{center}
\end{figure}

Polymers under flow are not only deformed, but also exhibit a
preferred, flow induced orientation
\cite{aust:99,schr:05,schr:05_1,wink:10,huan:10,huan:11}. To
characterize the degree of alignment, we calculate the probability
distribution of the angle $\phi$ between the end-to-end vector and
the flow direction. Examples are shown in Fig.~\ref{pdfang} for
various Weissenberg numbers and persistence lengths. There is no
preferred angle at equilibrium. With increasing shear rate, the
distribution function exhibits a maximum at a non-zero, positive
value $\phi_m$. This maximum shifts to smaller values with
increasing shear rate. At the same time the distribution function
becomes narrower. The latter implies that a polymer aligns
preferentially in a particular direction and samples other angles
only rarely.

Interestingly, the probability distribution functions become
asymmetric with increasing shear rate and exhibit a second maximum
at large angles. Thereby, the second maximum moves to larger
angles with increasing stiffness. The strong asymmetry, particular
for lower Weissenberg numbers, seems to be specific for two-dimensional
systems, because (flexible) polymers in three
dimensions \cite{huan:11} exhibit more symmetric distributions.
The second peak is peculiar for rather stiff polymers. However, it
is not clear whether it appears in two-dimensional systems only.
Here, studies of three-dimensional semiflexible polymers are
required to resolve the issue.
We report for completeness the fact that the appearance 
of two peaks in $P(\phi)$ at finite values, symmetric with respect
to $\phi=0$,  
was observed for two-dimensional grafted polymers with
the first bond fixed along the $x$-direction and $L_p \simeq L$ 
under equilibrium conditions.
\cite{latt:04}

\begin{figure}
\begin{center}
\includegraphics*[width=.6\textwidth,angle=0]{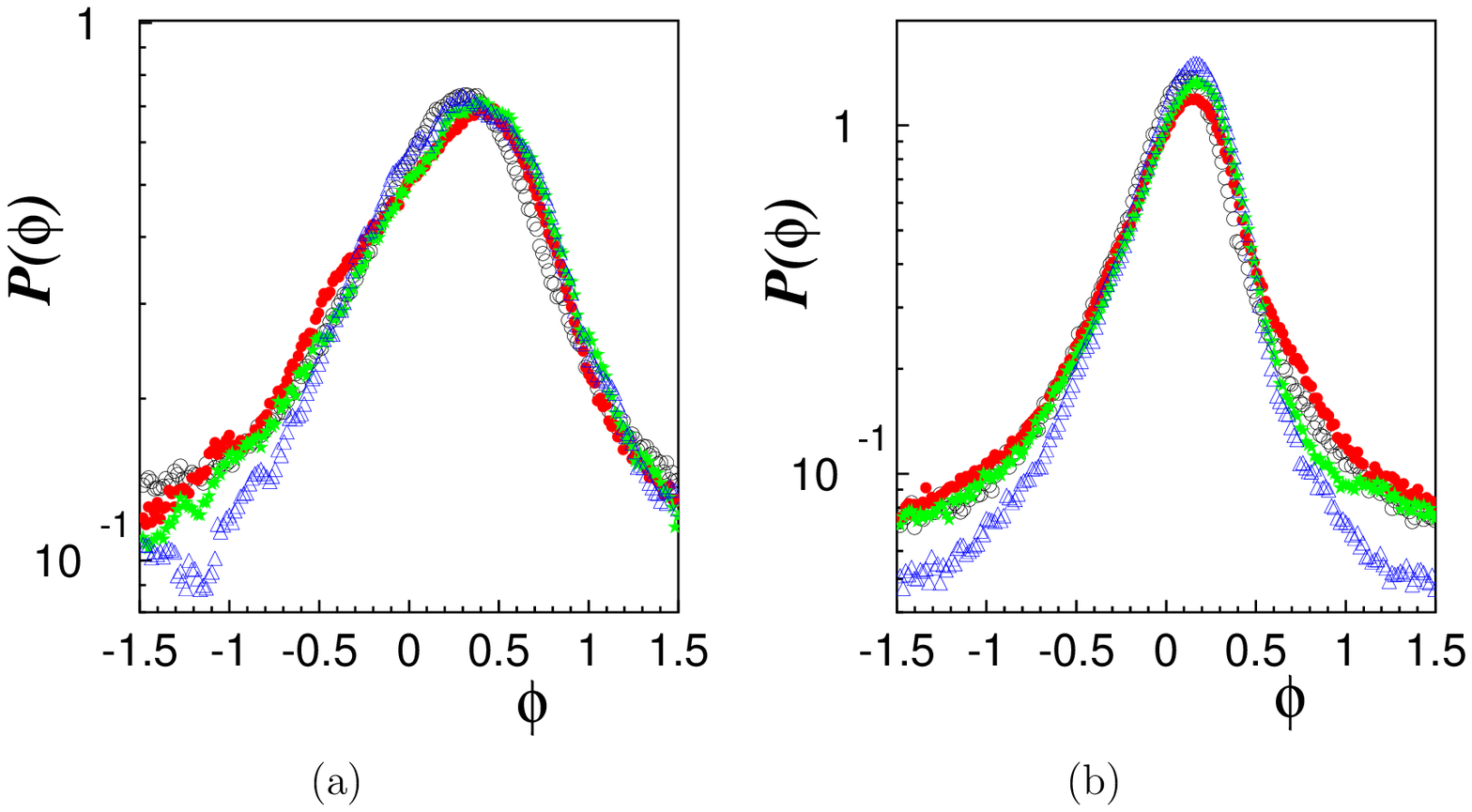}
\caption{
Probability distributions of the angle $\phi$
for the Weissenberg numbers $Wi
=8$ (a) and $ 80$ (b) and the
persistence lengths $L_p/L=0.1 (\circ), \ 0.4 ({\red \bullet}),
\ 2 ({\green \star}),$
and $10 ({\blue \triangle})$.
\label{pdfang_wi}}
\end{center}
\end{figure}

As shown in Fig.~\ref{pdfang_wi}, the probability distribution
functions depend only weakly on the persistence length for
Weissenberg numbers  $Wi \lesssim 10^2$. In particular, the
position of the maximum is virtually independent of $L_p$.

\begin{figure}[t]
\begin{center}
\includegraphics*[width=.45\textwidth,angle=0]{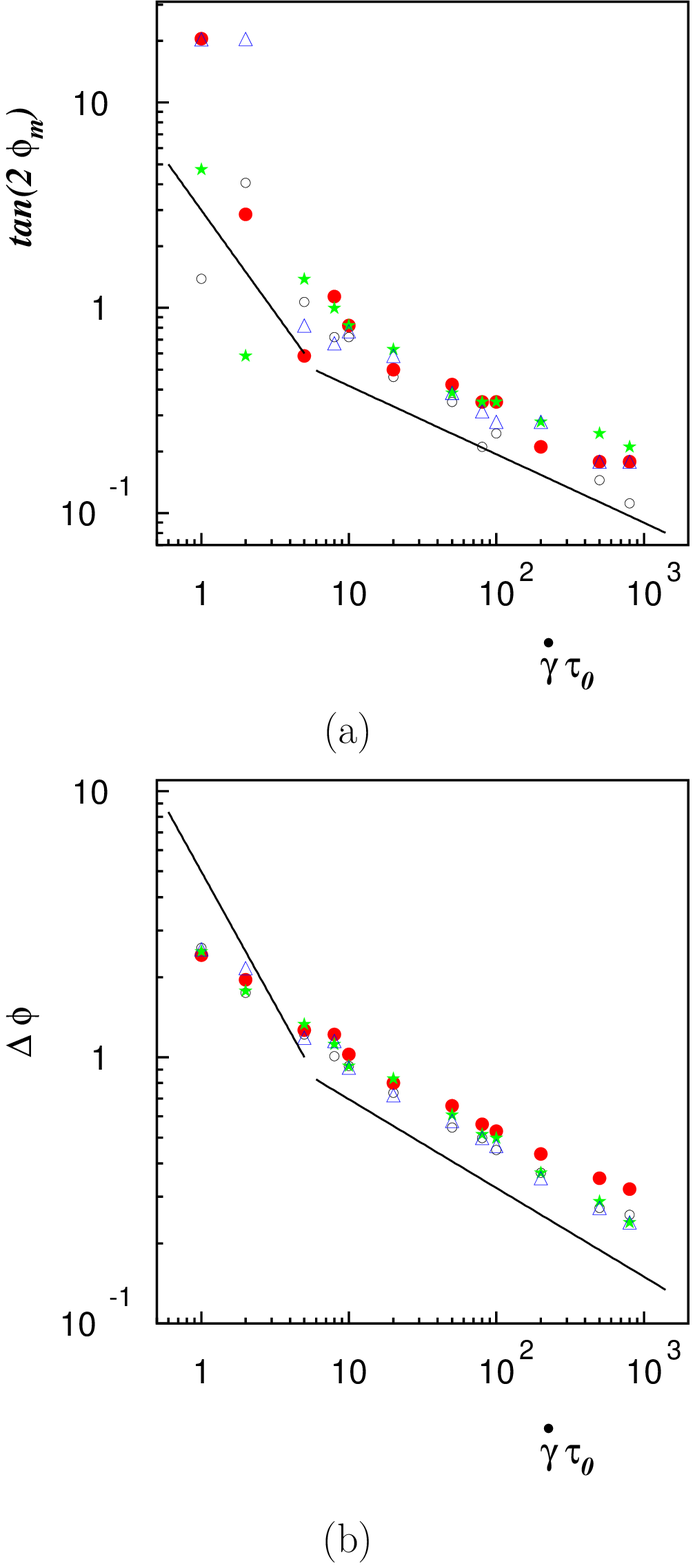}
\caption{(a) Angles $\phi_m$ of the maximum and (b) width $\Delta \phi$
of the distribution function $P(\phi)$ as function of the
Weissenberg number for the persistence lengths
$L_p/L=0.1 (\circ), \ 0.4 ({\red \bullet}), \ 2 ({\green \star}),$
and $10 ({\blue \triangle})$.
The slopes of the full lines are $-1$ and $-1/3$, respectively.
\label{tan}}
\end{center}
\end{figure}

Figure~\ref{tan} displays the angles $\phi_m$ of the central
maximum of $P(\phi)$. For small Weissenberg numbers $Wi <10$,
$\tan(2 \phi_m)$ decreases as  $Wi^{-1}$, whereas for larger
Weissenberg numbers the dependence $\tan(2 \phi_m) \sim Wi^{-1/3}$
is obtained. A similar dependence is found for flexible polymers
in three dimensions \cite{huan:11} and is predicted theoretically
\cite{wink:10,wink:06} independent of dimension. As already
suggested by Fig.~\ref{pdfang_wi}, nearly the same degree of
alignment is obtained independent of stiffnesses. However, the
values of $\tan(2 \phi_m)$ are somewhat smaller for the more
flexible polymers in the range $Wi>10$, as predicted theoretically
\cite{wink:10}.

Similarly, the width $\Delta \phi$, which is defined as the full
width at half maximum of the distribution function $P(\phi)$,
decreases as $\Delta \phi \sim Wi^{-1/3}$ at high shear rates.
This has also been observed in experiments \cite{gera:06} and
predicted theoretically \cite{wink:06}.

\section{Tumbling Dynamics} \label{sec:tumbling}

\begin{figure}
\begin{center}
\includegraphics*[width=.45\textwidth,angle=0]{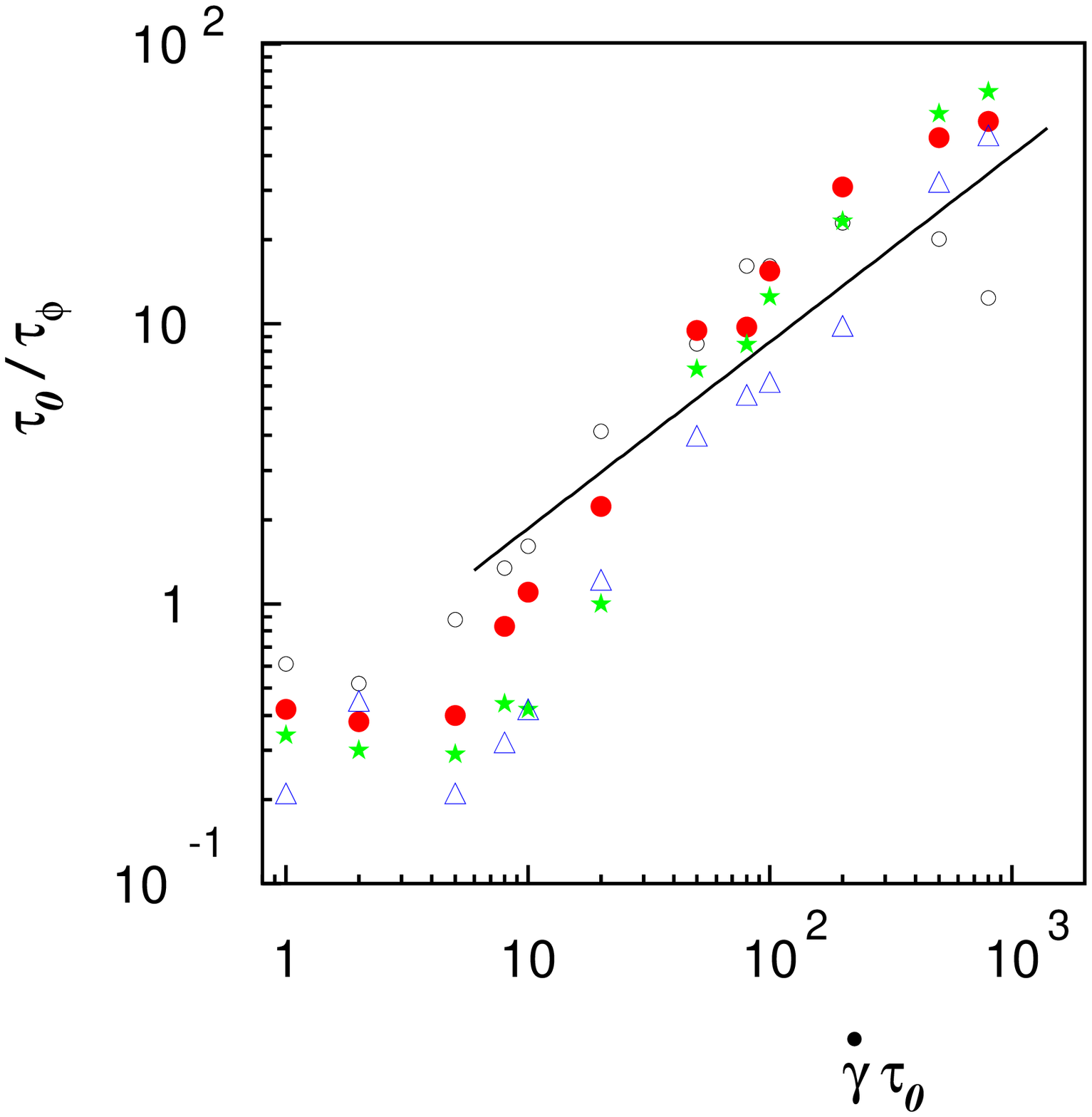}
\caption{Normalized tumbling frequencies $\tau_0/\tau_{\phi}$ as
function of the Weissenberg number for the persistence lengths
$L_p/L=0.1 (\circ), \ 0.4 ({\red \bullet}), \ 2 ({\green \star}),$ and
$10 ({\blue \triangle})$.
The line has the slope $2/3$. \label{relaxtau}}
\end{center}

\end{figure}

As mentioned in the Introduction, polymers in shear flow undergo a
tumbling motion. A characteristic tumbling time can be obtained
from the distribution function $P(t)$ of times between successive
zeros of the end-to-end vector component $R_{ex}(t)$ along the
flow direction.  This distribution exhibits the exponential decay
$P(t) \sim \exp{(-t/\tau_{\phi})}$ at large times, from which the
tumbling time $\tau_{\phi}$ is extracted. Normalized tumbling
frequencies  $\sim 1/\tau_{\phi}$  are depicted in
Fig.~\ref{relaxtau}, where the full line has slope $2/3$.

As for a three-dimensional system, we obtain $\tau_{\phi} \sim
Wi^{-2/3}$ for the shear rate dependence of the tumbling times.
This confirms that the tumbling times of semiflexible polymers
exhibit the same asymptotic Weissenberg number dependence as
flexible polymers. Moreover, two-dimensional tumbling behavior
seems to be similar to three-dimensional one for
semiflexible polymers.

Interestingly, the frequencies for the larger persistence lengths
seem to exhibit a fast, rather abrupt increase in the vicinity of
$Wi \approx 10$ and approach the dependence $Wi^{2/3}$ for large
Weissenberg numbers  only. Such a dependence has not been observed
in three dimensions so far nor be predicted theoretically. Whether
this is a specific two-dimensional features needs to be addressed
by simulations of semiflexible polymers in three dimensions.

\section{Conclusions} \label{sec:conclusion}

We have presented results for the non-equilibrium structural and
dynamical properties of semiflexible polymers confined to two
dimensions. The analysis of the force-extension relation of a
semiflexible polymer in a uniform external field yields excellent
agreement with theoretical predictions \cite{lamu:01.1,hori:07}
and confirms that the applied model is very well suited to
describe semiflexible polymers. Our studies of the end-to-end
distance relaxation behavior of initially stretched semiflexible
polymers confirm the experimentally obtained power-law time
dependence $t^{\gamma}$ with the exponent $\gamma = 0.45$, which
is in close agreement with the scaling prediction  $\gamma=1/2$ of
Ref.~\onlinecite{maie:02}.

We have also studied the conformation properties of semiflexible
polymers under shear flow. We clearly find strong shear-induced
conformational changes. Beyond a stiffness dependent Weissenberg
number, the average polymer extension decreases with increasing
shear rate, in contrast to flexible polymers, where the extension
increases. Visual inspection shows that U-shaped conformations
appear. Such conformations have also been observed for
semiflexible polymers in microchannel flows, both experimentally
\cite{stei:12} and in simulations \cite{chel:10,chel:11}. This
confirms that the end-to-end distances become similar in the
asymptotic  limit of infinite shear rate independent of the
stiffness \cite{wink:10}. As for flexible polymers in three
dimensions, the semiflexible polymers preferentially align along
the flow direction. However, the distribution functions of the
end-to-end vector alignment angle are clearly more asymmetric at
low Weissenberg numbers than for flexible polymers and exhibit a
second peak for large stiffnesses. It is not evident whether these
effects are caused by stiffness or confinement to two dimensions.
Further simulation studies are necessary to resolve this question.

We have also shown that semiflexible polymers exhibit a tumbling
motion, where the tumbling times approximately show the dependence
$\tau_{\phi} \sim Wi^{-2/3}$ on shear rate. Hence, semiflexible
polymers reveal in essence the same tumbling behavior as flexible
polymers
\cite{schr:05,schr:05_1,huan:11,huan:12,koba:10,wink:06_1,delg:06,wink:10}
and rods \cite{munk:06,wink:10}.

There are various aspects, e.g., the appearance of a second peak
in the orientational distribution functions and their asymmetry at
low shear rates, which need further investigations to clarify the
underlaying mechanism. This requires theoretical calculations
and/or simulations in two and three dimensions. We hope that our
results will stimulate such theoretical studies as well as
experimental investigations and will be valuable in the respective
endeavors.


\begin{thebibliography}{97}%
\makeatletter
\providecommand \@ifxundefined [1]{%
 \@ifx{#1\undefined}
}%
\providecommand \@ifnum [1]{%
 \ifnum #1\expandafter \@firstoftwo
 \else \expandafter \@secondoftwo
 \fi
}%
\providecommand \@ifx [1]{%
 \ifx #1\expandafter \@firstoftwo
 \else \expandafter \@secondoftwo
 \fi
}%
\providecommand \natexlab [1]{#1}%
\providecommand \enquote  [1]{``#1''}%
\providecommand \bibnamefont  [1]{#1}%
\providecommand \bibfnamefont [1]{#1}%
\providecommand \citenamefont [1]{#1}%
\providecommand \href@noop [0]{\@secondoftwo}%
\providecommand \href [0]{\begingroup \@sanitize@url \@href}%
\providecommand \@href[1]{\@@startlink{#1}\@@href}%
\providecommand \@@href[1]{\endgroup#1\@@endlink}%
\providecommand \@sanitize@url [0]{\catcode `\\12\catcode `\$12\catcode
  `\&12\catcode `\#12\catcode `\^12\catcode `\_12\catcode `\%12\relax}%
\providecommand \@@startlink[1]{}%
\providecommand \@@endlink[0]{}%
\providecommand \url  [0]{\begingroup\@sanitize@url \@url }%
\providecommand \@url [1]{\endgroup\@href {#1}{\urlprefix }}%
\providecommand \urlprefix  [0]{URL }%
\providecommand \Eprint [0]{\href }%
\providecommand \doibase [0]{http://dx.doi.org/}%
\providecommand \selectlanguage [0]{\@gobble}%
\providecommand \bibinfo  [0]{\@secondoftwo}%
\providecommand \bibfield  [0]{\@secondoftwo}%
\providecommand \translation [1]{[#1]}%
\providecommand \BibitemOpen [0]{}%
\providecommand \bibitemStop [0]{}%
\providecommand \bibitemNoStop [0]{.\EOS\space}%
\providecommand \EOS [0]{\spacefactor3000\relax}%
\providecommand \BibitemShut  [1]{\csname bibitem#1\endcsname}%
\let\auto@bib@innerbib\@empty
\bibitem [{\citenamefont {Janmey}(1995)}]{janm:95}%
  \BibitemOpen
  \bibfield  {author} {\bibinfo {author} {\bibfnamefont {P.}~\bibnamefont
  {Janmey}},\ }in\ \href@noop {} {\emph {\bibinfo {booktitle} {Handbook of
  Biological Physics}}},\ Vol.~\bibinfo {volume} {1A}\ (\bibinfo  {publisher}
  {North Holland},\ \bibinfo {address} {Amsterdam},\ \bibinfo {year}
  {1995})\BibitemShut {NoStop}%
\bibitem [{\citenamefont {Bustamante}\ \emph {et~al.}(1994)\citenamefont
  {Bustamante}, \citenamefont {Marko}, \citenamefont {Siggia},\ and\
  \citenamefont {Smith}}]{bust:94}%
  \BibitemOpen
  \bibfield  {author} {\bibinfo {author} {\bibfnamefont {C.}~\bibnamefont
  {Bustamante}}, \bibinfo {author} {\bibfnamefont {J.~F.}\ \bibnamefont
  {Marko}}, \bibinfo {author} {\bibfnamefont {E.~D.}\ \bibnamefont {Siggia}}, \
  and\ \bibinfo {author} {\bibfnamefont {S.}~\bibnamefont {Smith}},\
  }\href@noop {} {\bibfield  {journal} {\bibinfo  {journal} {Science}\ }\textbf
  {\bibinfo {volume} {265}},\ \bibinfo {pages} {1599} (\bibinfo {year}
  {1994})}\BibitemShut {NoStop}%
\bibitem [{\citenamefont {Marko}\ and\ \citenamefont {Siggia}(1995)}]{mark:95}%
  \BibitemOpen
  \bibfield  {author} {\bibinfo {author} {\bibfnamefont {J.~F.}\ \bibnamefont
  {Marko}}\ and\ \bibinfo {author} {\bibfnamefont {E.~D.}\ \bibnamefont
  {Siggia}},\ }\href@noop {} {\bibfield  {journal} {\bibinfo  {journal}
  {Macromolecules}\ }\textbf {\bibinfo {volume} {28}},\ \bibinfo {pages} {8759}
  (\bibinfo {year} {1995})}\BibitemShut {NoStop}%
\bibitem [{\citenamefont {Wilhelm}\ and\ \citenamefont {Frey}(1996)}]{wilh:96}%
  \BibitemOpen
  \bibfield  {author} {\bibinfo {author} {\bibfnamefont {J.}~\bibnamefont
  {Wilhelm}}\ and\ \bibinfo {author} {\bibfnamefont {E.}~\bibnamefont {Frey}},\
  }\href@noop {} {\bibfield  {journal} {\bibinfo  {journal} {Phys. Rev. Lett.}\
  }\textbf {\bibinfo {volume} {77}},\ \bibinfo {pages} {2581} (\bibinfo {year}
  {1996})}\BibitemShut {NoStop}%
\bibitem [{\citenamefont {Ripoll}\ \emph {et~al.}(2008)\citenamefont {Ripoll},
  \citenamefont {Holmqvist}, \citenamefont {Winkler}, \citenamefont {Gompper},
  \citenamefont {Dhont},\ and\ \citenamefont {Lettinga}}]{ripo:08}%
  \BibitemOpen
  \bibfield  {author} {\bibinfo {author} {\bibfnamefont {M.}~\bibnamefont
  {Ripoll}}, \bibinfo {author} {\bibfnamefont {P.}~\bibnamefont {Holmqvist}},
  \bibinfo {author} {\bibfnamefont {R.~G.}\ \bibnamefont {Winkler}}, \bibinfo
  {author} {\bibfnamefont {G.}~\bibnamefont {Gompper}}, \bibinfo {author}
  {\bibfnamefont {J.~K.~G.}\ \bibnamefont {Dhont}}, \ and\ \bibinfo {author}
  {\bibfnamefont {M.~P.}\ \bibnamefont {Lettinga}},\ }\href@noop {} {\bibfield
  {journal} {\bibinfo  {journal} {Phys. Rev. Lett.}\ }\textbf {\bibinfo
  {volume} {101}},\ \bibinfo {pages} {168302} (\bibinfo {year}
  {2008})}\BibitemShut {NoStop}%
\bibitem [{\citenamefont {Maeda}\ and\ \citenamefont {Fujime}(1984)}]{maed:84}%
  \BibitemOpen
  \bibfield  {author} {\bibinfo {author} {\bibfnamefont {T.}~\bibnamefont
  {Maeda}}\ and\ \bibinfo {author} {\bibfnamefont {S.}~\bibnamefont {Fujime}},\
  }\href@noop {} {\bibfield  {journal} {\bibinfo  {journal} {Macromolecules}\
  }\textbf {\bibinfo {volume} {17}},\ \bibinfo {pages} {2381} (\bibinfo {year}
  {1984})}\BibitemShut {NoStop}%
\bibitem [{\citenamefont {Arag\'{o}n}\ and\ \citenamefont
  {Pecora}(1985)}]{arag:85}%
  \BibitemOpen
  \bibfield  {author} {\bibinfo {author} {\bibfnamefont {S.~R.}\ \bibnamefont
  {Arag\'{o}n}}\ and\ \bibinfo {author} {\bibfnamefont {R.}~\bibnamefont
  {Pecora}},\ }\href@noop {} {\bibfield  {journal} {\bibinfo  {journal}
  {Macromolecules}\ }\textbf {\bibinfo {volume} {18}},\ \bibinfo {pages} {1868}
  (\bibinfo {year} {1985})}\BibitemShut {NoStop}%
\bibitem [{\citenamefont {Farge}\ and\ \citenamefont {Maggs}(1993)}]{farg:93}%
  \BibitemOpen
  \bibfield  {author} {\bibinfo {author} {\bibfnamefont {E.}~\bibnamefont
  {Farge}}\ and\ \bibinfo {author} {\bibfnamefont {A.~C.}\ \bibnamefont
  {Maggs}},\ }\href@noop {} {\bibfield  {journal} {\bibinfo  {journal}
  {Macromolecules}\ }\textbf {\bibinfo {volume} {26}},\ \bibinfo {pages} {5041}
  (\bibinfo {year} {1993})}\BibitemShut {NoStop}%
\bibitem [{\citenamefont {G{\"o}tter}\ \emph {et~al.}(1996)\citenamefont
  {G{\"o}tter}, \citenamefont {Kroy}, \citenamefont {Frey}, \citenamefont
  {B{\"a}rmann},\ and\ \citenamefont {Sackmann}}]{goet:96}%
  \BibitemOpen
  \bibfield  {author} {\bibinfo {author} {\bibfnamefont {R.}~\bibnamefont
  {G{\"o}tter}}, \bibinfo {author} {\bibfnamefont {K.}~\bibnamefont {Kroy}},
  \bibinfo {author} {\bibfnamefont {E.}~\bibnamefont {Frey}}, \bibinfo {author}
  {\bibfnamefont {M.}~\bibnamefont {B{\"a}rmann}}, \ and\ \bibinfo {author}
  {\bibfnamefont {E.}~\bibnamefont {Sackmann}},\ }\href@noop {} {\bibfield
  {journal} {\bibinfo  {journal} {Macromolecules}\ }\textbf {\bibinfo {volume}
  {29}},\ \bibinfo {pages} {30} (\bibinfo {year} {1996})}\BibitemShut {NoStop}%
\bibitem [{\citenamefont {Harnau}, \citenamefont {Winkler},\ and\ \citenamefont
  {Reineker}(1996)}]{harn:96}%
  \BibitemOpen
  \bibfield  {author} {\bibinfo {author} {\bibfnamefont {L.}~\bibnamefont
  {Harnau}}, \bibinfo {author} {\bibfnamefont {R.~G.}\ \bibnamefont {Winkler}},
  \ and\ \bibinfo {author} {\bibfnamefont {P.}~\bibnamefont {Reineker}},\
  }\href@noop {} {\bibfield  {journal} {\bibinfo  {journal} {J. Chem. Phys.}\
  }\textbf {\bibinfo {volume} {104}},\ \bibinfo {pages} {6355} (\bibinfo {year}
  {1996})}\BibitemShut {NoStop}%
\bibitem [{\citenamefont {Everaers}\ \emph {et~al.}(1999)\citenamefont
  {Everaers}, \citenamefont {J\"ulicher}, \citenamefont {Ajdari},\ and\
  \citenamefont {Maggs}}]{ever:99}%
  \BibitemOpen
  \bibfield  {author} {\bibinfo {author} {\bibfnamefont {R.}~\bibnamefont
  {Everaers}}, \bibinfo {author} {\bibfnamefont {F.}~\bibnamefont
  {J\"ulicher}}, \bibinfo {author} {\bibfnamefont {A.}~\bibnamefont {Ajdari}},
  \ and\ \bibinfo {author} {\bibfnamefont {A.~C.}\ \bibnamefont {Maggs}},\
  }\href@noop {} {\bibfield  {journal} {\bibinfo  {journal} {Phys. Rev. Lett.}\
  }\textbf {\bibinfo {volume} {82}},\ \bibinfo {pages} {3717} (\bibinfo {year}
  {1999})}\BibitemShut {NoStop}%
\bibitem [{\citenamefont {Samuel}\ and\ \citenamefont {Sinha}(2002)}]{samu:02}%
  \BibitemOpen
  \bibfield  {author} {\bibinfo {author} {\bibfnamefont {J.}~\bibnamefont
  {Samuel}}\ and\ \bibinfo {author} {\bibfnamefont {S.}~\bibnamefont {Sinha}},\
  }\href@noop {} {\bibfield  {journal} {\bibinfo  {journal} {Phys. Rev. E}\
  }\textbf {\bibinfo {volume} {66}},\ \bibinfo {pages} {050801} (\bibinfo
  {year} {2002})}\BibitemShut {NoStop}%
\bibitem [{\citenamefont {Le~Goff}\ \emph {et~al.}(2002)\citenamefont
  {Le~Goff}, \citenamefont {Hallatschek}, \citenamefont {Frey},\ and\
  \citenamefont {Amblard}}]{lego:02}%
  \BibitemOpen
  \bibfield  {author} {\bibinfo {author} {\bibfnamefont {L.}~\bibnamefont
  {Le~Goff}}, \bibinfo {author} {\bibfnamefont {O.}~\bibnamefont
  {Hallatschek}}, \bibinfo {author} {\bibfnamefont {E.}~\bibnamefont {Frey}}, \
  and\ \bibinfo {author} {\bibfnamefont {F.}~\bibnamefont {Amblard}},\
  }\href@noop {} {\bibfield  {journal} {\bibinfo  {journal} {Phys. Rev. Lett.}\
  }\textbf {\bibinfo {volume} {89}},\ \bibinfo {pages} {258101} (\bibinfo
  {year} {2002})}\BibitemShut {NoStop}%
\bibitem [{\citenamefont {Winkler}(2003)}]{wink:03}%
  \BibitemOpen
  \bibfield  {author} {\bibinfo {author} {\bibfnamefont {R.~G.}\ \bibnamefont
  {Winkler}},\ }\href@noop {} {\bibfield  {journal} {\bibinfo  {journal} {J.
  Chem. Phys.}\ }\textbf {\bibinfo {volume} {118}},\ \bibinfo {pages} {2919}
  (\bibinfo {year} {2003})}\BibitemShut {NoStop}%
\bibitem [{\citenamefont {Carria}(2004)}]{carr:04}%
  \BibitemOpen
  \bibfield  {author} {\bibinfo {author} {\bibfnamefont {G.~A.}\ \bibnamefont
  {Carria}},\ }\href@noop {} {\bibfield  {journal} {\bibinfo  {journal} {J.
  Chem. Phys.}\ }\textbf {\bibinfo {volume} {121}},\ \bibinfo {pages} {12112}
  (\bibinfo {year} {2004})}\BibitemShut {NoStop}%
\bibitem [{\citenamefont {Petrov}\ \emph {et~al.}(2006)\citenamefont {Petrov},
  \citenamefont {Ohrt}, \citenamefont {Winkler},\ and\ \citenamefont
  {Schwille}}]{petr:06}%
  \BibitemOpen
  \bibfield  {author} {\bibinfo {author} {\bibfnamefont {E.~P.}\ \bibnamefont
  {Petrov}}, \bibinfo {author} {\bibfnamefont {T.}~\bibnamefont {Ohrt}},
  \bibinfo {author} {\bibfnamefont {R.~G.}\ \bibnamefont {Winkler}}, \ and\
  \bibinfo {author} {\bibfnamefont {P.}~\bibnamefont {Schwille}},\ }\href@noop
  {} {\bibfield  {journal} {\bibinfo  {journal} {Phys. Rev. Lett.}\ }\textbf
  {\bibinfo {volume} {97}},\ \bibinfo {pages} {258101} (\bibinfo {year}
  {2006})}\BibitemShut {NoStop}%
\bibitem [{\citenamefont {Winkler}, \citenamefont {Keller},\ and\ \citenamefont
  {R{\"a}dler}(2006)}]{wink:06}%
  \BibitemOpen
  \bibfield  {author} {\bibinfo {author} {\bibfnamefont {R.~G.}\ \bibnamefont
  {Winkler}}, \bibinfo {author} {\bibfnamefont {S.}~\bibnamefont {Keller}}, \
  and\ \bibinfo {author} {\bibfnamefont {J.~O.}\ \bibnamefont {R{\"a}dler}},\
  }\href@noop {} {\bibfield  {journal} {\bibinfo  {journal} {Phys. Rev. E}\
  }\textbf {\bibinfo {volume} {73}},\ \bibinfo {pages} {041919} (\bibinfo
  {year} {2006})}\BibitemShut {NoStop}%
\bibitem [{\citenamefont {Winkler}(2007)}]{wink:07.1}%
  \BibitemOpen
  \bibfield  {author} {\bibinfo {author} {\bibfnamefont {R.~G.}\ \bibnamefont
  {Winkler}},\ }\href@noop {} {\bibfield  {journal} {\bibinfo  {journal} {J.
  Chem. Phys.}\ }\textbf {\bibinfo {volume} {127}},\ \bibinfo {pages} {054904}
  (\bibinfo {year} {2007})}\BibitemShut {NoStop}%
\bibitem [{\citenamefont {Perkins}\ \emph {et~al.}(1995)\citenamefont
  {Perkins}, \citenamefont {Smith}, \citenamefont {Larson},\ and\ \citenamefont
  {Chu}}]{perk:95}%
  \BibitemOpen
  \bibfield  {author} {\bibinfo {author} {\bibfnamefont {T.~T.}\ \bibnamefont
  {Perkins}}, \bibinfo {author} {\bibfnamefont {D.~E.}\ \bibnamefont {Smith}},
  \bibinfo {author} {\bibfnamefont {R.~G.}\ \bibnamefont {Larson}}, \ and\
  \bibinfo {author} {\bibfnamefont {S.}~\bibnamefont {Chu}},\ }\href@noop {}
  {\bibfield  {journal} {\bibinfo  {journal} {Science}\ }\textbf {\bibinfo
  {volume} {268}},\ \bibinfo {pages} {83} (\bibinfo {year} {1995})}\BibitemShut
  {NoStop}%
\bibitem [{\citenamefont {Perkins}, \citenamefont {Smith},\ and\ \citenamefont
  {Chu}(1997)}]{perk:97}%
  \BibitemOpen
  \bibfield  {author} {\bibinfo {author} {\bibfnamefont {T.~T.}\ \bibnamefont
  {Perkins}}, \bibinfo {author} {\bibfnamefont {D.~E.}\ \bibnamefont {Smith}},
  \ and\ \bibinfo {author} {\bibfnamefont {S.}~\bibnamefont {Chu}},\
  }\href@noop {} {\bibfield  {journal} {\bibinfo  {journal} {Science}\ }\textbf
  {\bibinfo {volume} {276}},\ \bibinfo {pages} {2016} (\bibinfo {year}
  {1997})}\BibitemShut {NoStop}%
\bibitem [{\citenamefont {Quake}, \citenamefont {Babcock},\ and\ \citenamefont
  {Chu}(1997)}]{quak:97}%
  \BibitemOpen
  \bibfield  {author} {\bibinfo {author} {\bibfnamefont {S.~R.}\ \bibnamefont
  {Quake}}, \bibinfo {author} {\bibfnamefont {H.}~\bibnamefont {Babcock}}, \
  and\ \bibinfo {author} {\bibfnamefont {S.}~\bibnamefont {Chu}},\ }\href@noop
  {} {\bibfield  {journal} {\bibinfo  {journal} {Nature}\ }\textbf {\bibinfo
  {volume} {388}},\ \bibinfo {pages} {151} (\bibinfo {year}
  {1997})}\BibitemShut {NoStop}%
\bibitem [{\citenamefont {Winkler}(1999)}]{wink:99}%
  \BibitemOpen
  \bibfield  {author} {\bibinfo {author} {\bibfnamefont {R.~G.}\ \bibnamefont
  {Winkler}},\ }\href@noop {} {\bibfield  {journal} {\bibinfo  {journal} {Phys.
  Rev. Lett.}\ }\textbf {\bibinfo {volume} {82}},\ \bibinfo {pages} {1843}
  (\bibinfo {year} {1999})}\BibitemShut {NoStop}%
\bibitem [{\citenamefont {LeDuc}\ \emph {et~al.}(1999)\citenamefont {LeDuc},
  \citenamefont {Haber}, \citenamefont {Boa},\ and\ \citenamefont
  {Wirtz}}]{ledu:99}%
  \BibitemOpen
  \bibfield  {author} {\bibinfo {author} {\bibfnamefont {P.}~\bibnamefont
  {LeDuc}}, \bibinfo {author} {\bibfnamefont {C.}~\bibnamefont {Haber}},
  \bibinfo {author} {\bibfnamefont {G.}~\bibnamefont {Boa}}, \ and\ \bibinfo
  {author} {\bibfnamefont {D.}~\bibnamefont {Wirtz}},\ }\href@noop {}
  {\bibfield  {journal} {\bibinfo  {journal} {Nature}\ }\textbf {\bibinfo
  {volume} {399}},\ \bibinfo {pages} {564} (\bibinfo {year}
  {1999})}\BibitemShut {NoStop}%
\bibitem [{\citenamefont {Smith}, \citenamefont {Babcock},\ and\ \citenamefont
  {Chu}(1999)}]{smit:99}%
  \BibitemOpen
  \bibfield  {author} {\bibinfo {author} {\bibfnamefont {D.~E.}\ \bibnamefont
  {Smith}}, \bibinfo {author} {\bibfnamefont {H.~P.}\ \bibnamefont {Babcock}},
  \ and\ \bibinfo {author} {\bibfnamefont {S.}~\bibnamefont {Chu}},\
  }\href@noop {} {\bibfield  {journal} {\bibinfo  {journal} {Science}\ }\textbf
  {\bibinfo {volume} {283}},\ \bibinfo {pages} {1724} (\bibinfo {year}
  {1999})}\BibitemShut {NoStop}%
\bibitem [{\citenamefont {Schroeder}\ \emph
  {et~al.}(2005{\natexlab{a}})\citenamefont {Schroeder}, \citenamefont
  {Teixeira}, \citenamefont {Shaqfeh},\ and\ \citenamefont {Chu}}]{schr:05}%
  \BibitemOpen
  \bibfield  {author} {\bibinfo {author} {\bibfnamefont {C.~M.}\ \bibnamefont
  {Schroeder}}, \bibinfo {author} {\bibfnamefont {R.~E.}\ \bibnamefont
  {Teixeira}}, \bibinfo {author} {\bibfnamefont {E.~S.~G.}\ \bibnamefont
  {Shaqfeh}}, \ and\ \bibinfo {author} {\bibfnamefont {S.}~\bibnamefont
  {Chu}},\ }\href@noop {} {\bibfield  {journal} {\bibinfo  {journal} {Phys.
  Rev. Lett.}\ }\textbf {\bibinfo {volume} {95}},\ \bibinfo {pages} {018301}
  (\bibinfo {year} {2005}{\natexlab{a}})}\BibitemShut {NoStop}%
\bibitem [{\citenamefont {Gerashchenko}\ and\ \citenamefont
  {Steinberg}(2006)}]{gera:06}%
  \BibitemOpen
  \bibfield  {author} {\bibinfo {author} {\bibfnamefont {S.}~\bibnamefont
  {Gerashchenko}}\ and\ \bibinfo {author} {\bibfnamefont {V.}~\bibnamefont
  {Steinberg}},\ }\href@noop {} {\bibfield  {journal} {\bibinfo  {journal}
  {Phys. Rev. Lett.}\ }\textbf {\bibinfo {volume} {96}},\ \bibinfo {pages}
  {038304} (\bibinfo {year} {2006})}\BibitemShut {NoStop}%
\bibitem [{\citenamefont {Teixeira}\ \emph {et~al.}(2005)\citenamefont
  {Teixeira}, \citenamefont {Babcock}, \citenamefont {Shaqfeh},\ and\
  \citenamefont {Chu}}]{teix:05}%
  \BibitemOpen
  \bibfield  {author} {\bibinfo {author} {\bibfnamefont {R.~E.}\ \bibnamefont
  {Teixeira}}, \bibinfo {author} {\bibfnamefont {H.~P.}\ \bibnamefont
  {Babcock}}, \bibinfo {author} {\bibfnamefont {E.~S.~G.}\ \bibnamefont
  {Shaqfeh}}, \ and\ \bibinfo {author} {\bibfnamefont {S.}~\bibnamefont
  {Chu}},\ }\href@noop {} {\bibfield  {journal} {\bibinfo  {journal}
  {Macromolecules}\ }\textbf {\bibinfo {volume} {38}},\ \bibinfo {pages} {581}
  (\bibinfo {year} {2005})}\BibitemShut {NoStop}%
\bibitem [{\citenamefont {Schroeder}\ \emph
  {et~al.}(2005{\natexlab{b}})\citenamefont {Schroeder}, \citenamefont
  {Teixeira}, \citenamefont {Shaqfeh},\ and\ \citenamefont {Chu}}]{schr:05_1}%
  \BibitemOpen
  \bibfield  {author} {\bibinfo {author} {\bibfnamefont {C.~M.}\ \bibnamefont
  {Schroeder}}, \bibinfo {author} {\bibfnamefont {R.~E.}\ \bibnamefont
  {Teixeira}}, \bibinfo {author} {\bibfnamefont {E.~S.~G.}\ \bibnamefont
  {Shaqfeh}}, \ and\ \bibinfo {author} {\bibfnamefont {S.}~\bibnamefont
  {Chu}},\ }\href@noop {} {\bibfield  {journal} {\bibinfo  {journal}
  {Macromolecules}\ }\textbf {\bibinfo {volume} {38}},\ \bibinfo {pages} {1967}
  (\bibinfo {year} {2005}{\natexlab{b}})}\BibitemShut {NoStop}%
\bibitem [{\citenamefont {Doyle}, \citenamefont {Ladoux},\ and\ \citenamefont
  {Viovy}(2000)}]{doyl:00}%
  \BibitemOpen
  \bibfield  {author} {\bibinfo {author} {\bibfnamefont {P.~S.}\ \bibnamefont
  {Doyle}}, \bibinfo {author} {\bibfnamefont {B.}~\bibnamefont {Ladoux}}, \
  and\ \bibinfo {author} {\bibfnamefont {J.-L.}\ \bibnamefont {Viovy}},\
  }\href@noop {} {\bibfield  {journal} {\bibinfo  {journal} {Phys. Rev. Lett.}\
  }\textbf {\bibinfo {volume} {84}},\ \bibinfo {pages} {4769} (\bibinfo {year}
  {2000})}\BibitemShut {NoStop}%
\bibitem [{\citenamefont {Ladoux}\ and\ \citenamefont {Doyle}(2000)}]{lado:00}%
  \BibitemOpen
  \bibfield  {author} {\bibinfo {author} {\bibfnamefont {B.}~\bibnamefont
  {Ladoux}}\ and\ \bibinfo {author} {\bibfnamefont {P.~S.}\ \bibnamefont
  {Doyle}},\ }\href@noop {} {\bibfield  {journal} {\bibinfo  {journal}
  {Europhys. Lett.}\ }\textbf {\bibinfo {volume} {52}},\ \bibinfo {pages} {511}
  (\bibinfo {year} {2000})}\BibitemShut {NoStop}%
\bibitem [{\citenamefont {Johnson}(1987)}]{john:87}%
  \BibitemOpen
  \bibfield  {author} {\bibinfo {author} {\bibfnamefont {J.~A.~Y.}\
  \bibnamefont {Johnson}},\ }\href@noop {} {\bibfield  {journal} {\bibinfo
  {journal} {Macromolecules}\ }\textbf {\bibinfo {volume} {20}},\ \bibinfo
  {pages} {103} (\bibinfo {year} {1987})}\BibitemShut {NoStop}%
\bibitem [{\citenamefont {Bruns}\ and\ \citenamefont {Carl}(1993)}]{brun:93}%
  \BibitemOpen
  \bibfield  {author} {\bibinfo {author} {\bibfnamefont {W.}~\bibnamefont
  {Bruns}}\ and\ \bibinfo {author} {\bibfnamefont {W.}~\bibnamefont {Carl}},\
  }\href@noop {} {\bibfield  {journal} {\bibinfo  {journal} {Macromolecules}\
  }\textbf {\bibinfo {volume} {26}},\ \bibinfo {pages} {557} (\bibinfo {year}
  {1993})}\BibitemShut {NoStop}%
\bibitem [{\citenamefont {Carl}\ and\ \citenamefont {Bruns}(1994)}]{carl:94}%
  \BibitemOpen
  \bibfield  {author} {\bibinfo {author} {\bibfnamefont {W.}~\bibnamefont
  {Carl}}\ and\ \bibinfo {author} {\bibfnamefont {W.}~\bibnamefont {Bruns}},\
  }\href@noop {} {\bibfield  {journal} {\bibinfo  {journal} {Macromol. Theory
  Simul.}\ }\textbf {\bibinfo {volume} {3}},\ \bibinfo {pages} {295} (\bibinfo
  {year} {1994})}\BibitemShut {NoStop}%
\bibitem [{\citenamefont {Wang}\ and\ \citenamefont
  {Chatterjee}(2001)}]{wang:01}%
  \BibitemOpen
  \bibfield  {author} {\bibinfo {author} {\bibfnamefont {X.}~\bibnamefont
  {Wang}}\ and\ \bibinfo {author} {\bibfnamefont {A.~P.}\ \bibnamefont
  {Chatterjee}},\ }\href@noop {} {\bibfield  {journal} {\bibinfo  {journal}
  {Macromolecules}\ }\textbf {\bibinfo {volume} {34}},\ \bibinfo {pages} {1118}
  (\bibinfo {year} {2001})}\BibitemShut {NoStop}%
\bibitem [{\citenamefont {Woo}\ and\ \citenamefont {Shaqfeh}(0030)}]{woo:03}%
  \BibitemOpen
  \bibfield  {author} {\bibinfo {author} {\bibfnamefont {N.~J.}\ \bibnamefont
  {Woo}}\ and\ \bibinfo {author} {\bibfnamefont {E.~S.~G.}\ \bibnamefont
  {Shaqfeh}},\ }\href@noop {} {\bibfield  {journal} {\bibinfo  {journal} {J.
  Chem. Phys.}\ }\textbf {\bibinfo {volume} {119}},\ \bibinfo {pages} {2908}
  (\bibinfo {year} {20030})}\BibitemShut {NoStop}%
\bibitem [{\citenamefont {Dubbeldam}\ and\ \citenamefont
  {Redig}(2006)}]{dubb:06}%
  \BibitemOpen
  \bibfield  {author} {\bibinfo {author} {\bibfnamefont {J.}~\bibnamefont
  {Dubbeldam}}\ and\ \bibinfo {author} {\bibfnamefont {F.}~\bibnamefont
  {Redig}},\ }\href@noop {} {\bibfield  {journal} {\bibinfo  {journal} {J.
  Stat. Phys.}\ }\textbf {\bibinfo {volume} {125}},\ \bibinfo {pages} {225}
  (\bibinfo {year} {2006})}\BibitemShut {NoStop}%
\bibitem [{\citenamefont {Bird}\ \emph {et~al.}(1987)\citenamefont {Bird},
  \citenamefont {Curtiss}, \citenamefont {Armstrong},\ and\ \citenamefont
  {Hassager}}]{bird:87}%
  \BibitemOpen
  \bibfield  {author} {\bibinfo {author} {\bibfnamefont {R.~B.}\ \bibnamefont
  {Bird}}, \bibinfo {author} {\bibfnamefont {C.~F.}\ \bibnamefont {Curtiss}},
  \bibinfo {author} {\bibfnamefont {R.~C.}\ \bibnamefont {Armstrong}}, \ and\
  \bibinfo {author} {\bibfnamefont {O.}~\bibnamefont {Hassager}},\ }\href@noop
  {} {\emph {\bibinfo {title} {Dynamics of Polymer Liquids}}},\ Vol.~\bibinfo
  {volume} {2}\ (\bibinfo  {publisher} {John Wiley \& Sons},\ \bibinfo
  {address} {New York},\ \bibinfo {year} {1987})\BibitemShut {NoStop}%
\bibitem [{\citenamefont {{\"O}ttinger}(1996)}]{oett:96}%
  \BibitemOpen
  \bibfield  {author} {\bibinfo {author} {\bibfnamefont {H.~C.}\ \bibnamefont
  {{\"O}ttinger}},\ }\href@noop {} {\emph {\bibinfo {title} {Stochastic
  Processes in Polymeric Fluids}}}\ (\bibinfo  {publisher} {Springer},\
  \bibinfo {address} {Berlin},\ \bibinfo {year} {1996})\BibitemShut {NoStop}%
\bibitem [{\citenamefont {Dua}\ and\ \citenamefont {Cherayil}(2000)}]{dua:00}%
  \BibitemOpen
  \bibfield  {author} {\bibinfo {author} {\bibfnamefont {A.}~\bibnamefont
  {Dua}}\ and\ \bibinfo {author} {\bibfnamefont {B.~J.}\ \bibnamefont
  {Cherayil}},\ }\href@noop {} {\bibfield  {journal} {\bibinfo  {journal} {J.
  Chem. Phys.}\ }\textbf {\bibinfo {volume} {112}},\ \bibinfo {pages} {8707}
  (\bibinfo {year} {2000})}\BibitemShut {NoStop}%
\bibitem [{\citenamefont {Rabin}\ and\ \citenamefont
  {Kawasaki}(1989)}]{rabi:89}%
  \BibitemOpen
  \bibfield  {author} {\bibinfo {author} {\bibfnamefont {Y.}~\bibnamefont
  {Rabin}}\ and\ \bibinfo {author} {\bibfnamefont {K.}~\bibnamefont
  {Kawasaki}},\ }\href@noop {} {\bibfield  {journal} {\bibinfo  {journal}
  {Phys. Rev. Lett.}\ }\textbf {\bibinfo {volume} {62}},\ \bibinfo {pages}
  {2281} (\bibinfo {year} {1989})}\BibitemShut {NoStop}%
\bibitem [{\citenamefont {Prakash}(2001)}]{prak:01}%
  \BibitemOpen
  \bibfield  {author} {\bibinfo {author} {\bibfnamefont {J.~R.}\ \bibnamefont
  {Prakash}},\ }\href@noop {} {\bibfield  {journal} {\bibinfo  {journal}
  {Macromolecules}\ }\textbf {\bibinfo {volume} {34}},\ \bibinfo {pages} {3396}
  (\bibinfo {year} {2001})}\BibitemShut {NoStop}%
\bibitem [{\citenamefont {Wang}(1989)}]{wang:89}%
  \BibitemOpen
  \bibfield  {author} {\bibinfo {author} {\bibfnamefont {S.-Q.}\ \bibnamefont
  {Wang}},\ }\href@noop {} {\bibfield  {journal} {\bibinfo  {journal} {Phys.
  Rev. A}\ }\textbf {\bibinfo {volume} {40}},\ \bibinfo {pages} {2137}
  (\bibinfo {year} {1989})}\BibitemShut {NoStop}%
\bibitem [{\citenamefont {Baldwin}\ and\ \citenamefont
  {Helfand}(1990)}]{bald:90}%
  \BibitemOpen
  \bibfield  {author} {\bibinfo {author} {\bibfnamefont {P.~R.}\ \bibnamefont
  {Baldwin}}\ and\ \bibinfo {author} {\bibfnamefont {E.}~\bibnamefont
  {Helfand}},\ }\href@noop {} {\bibfield  {journal} {\bibinfo  {journal} {Phys.
  Rev. A}\ }\textbf {\bibinfo {volume} {41}},\ \bibinfo {pages} {6772}
  (\bibinfo {year} {1990})}\BibitemShut {NoStop}%
\bibitem [{\citenamefont {Ganazzoli}\ and\ \citenamefont
  {Raffaini}(1999)}]{gana:99}%
  \BibitemOpen
  \bibfield  {author} {\bibinfo {author} {\bibfnamefont {F.}~\bibnamefont
  {Ganazzoli}}\ and\ \bibinfo {author} {\bibfnamefont {G.}~\bibnamefont
  {Raffaini}},\ }\href@noop {} {\bibfield  {journal} {\bibinfo  {journal}
  {Macromol. Theory Simul.}\ }\textbf {\bibinfo {volume} {8}},\ \bibinfo
  {pages} {234} (\bibinfo {year} {1999})}\BibitemShut {NoStop}%
\bibitem [{\citenamefont {Subbotin}\ \emph {et~al.}(1995)\citenamefont
  {Subbotin}, \citenamefont {Semenov}, \citenamefont {Manias}, \citenamefont
  {Hadziioannou},\ and\ \citenamefont {ten Brinke}}]{subb:95}%
  \BibitemOpen
  \bibfield  {author} {\bibinfo {author} {\bibfnamefont {A.}~\bibnamefont
  {Subbotin}}, \bibinfo {author} {\bibfnamefont {A.}~\bibnamefont {Semenov}},
  \bibinfo {author} {\bibfnamefont {E.}~\bibnamefont {Manias}}, \bibinfo
  {author} {\bibfnamefont {G.}~\bibnamefont {Hadziioannou}}, \ and\ \bibinfo
  {author} {\bibfnamefont {G.}~\bibnamefont {ten Brinke}},\ }\href@noop {}
  {\bibfield  {journal} {\bibinfo  {journal} {Macromolecules}\ }\textbf
  {\bibinfo {volume} {28}},\ \bibinfo {pages} {3898} (\bibinfo {year}
  {1995})}\BibitemShut {NoStop}%
\bibitem [{\citenamefont {Prabhakar}\ and\ \citenamefont
  {Prakash}(2006)}]{prab:06}%
  \BibitemOpen
  \bibfield  {author} {\bibinfo {author} {\bibfnamefont {R.}~\bibnamefont
  {Prabhakar}}\ and\ \bibinfo {author} {\bibfnamefont {J.~R.}\ \bibnamefont
  {Prakash}},\ }\href@noop {} {\bibfield  {journal} {\bibinfo  {journal} {J.
  Rheol.}\ }\textbf {\bibinfo {volume} {50}},\ \bibinfo {pages} {561} (\bibinfo
  {year} {2006})}\BibitemShut {NoStop}%
\bibitem [{\citenamefont {Winkler}(2006)}]{wink:06_1}%
  \BibitemOpen
  \bibfield  {author} {\bibinfo {author} {\bibfnamefont {R.~G.}\ \bibnamefont
  {Winkler}},\ }\href@noop {} {\bibfield  {journal} {\bibinfo  {journal} {Phys.
  Rev. Lett.}\ }\textbf {\bibinfo {volume} {97}},\ \bibinfo {pages} {128301}
  (\bibinfo {year} {2006})}\BibitemShut {NoStop}%
\bibitem [{\citenamefont {Munk}\ \emph {et~al.}(2006)\citenamefont {Munk},
  \citenamefont {Hallatschek}, \citenamefont {Wiggins},\ and\ \citenamefont
  {Frey}}]{munk:06}%
  \BibitemOpen
  \bibfield  {author} {\bibinfo {author} {\bibfnamefont {T.}~\bibnamefont
  {Munk}}, \bibinfo {author} {\bibfnamefont {O.}~\bibnamefont {Hallatschek}},
  \bibinfo {author} {\bibfnamefont {C.~H.}\ \bibnamefont {Wiggins}}, \ and\
  \bibinfo {author} {\bibfnamefont {E.}~\bibnamefont {Frey}},\ }\href@noop {}
  {\bibfield  {journal} {\bibinfo  {journal} {Phys. Rev. E}\ }\textbf {\bibinfo
  {volume} {74}},\ \bibinfo {pages} {041911} (\bibinfo {year}
  {2006})}\BibitemShut {NoStop}%
\bibitem [{\citenamefont {Winkler}(2010)}]{wink:10}%
  \BibitemOpen
  \bibfield  {author} {\bibinfo {author} {\bibfnamefont {R.~G.}\ \bibnamefont
  {Winkler}},\ }\href@noop {} {\bibfield  {journal} {\bibinfo  {journal} {J.
  Chem. Phys.}\ }\textbf {\bibinfo {volume} {133}},\ \bibinfo {pages} {164905}
  (\bibinfo {year} {2010})}\BibitemShut {NoStop}%
\bibitem [{\citenamefont {Liu}(1989)}]{liu:89}%
  \BibitemOpen
  \bibfield  {author} {\bibinfo {author} {\bibfnamefont {T.}~\bibnamefont
  {Liu}},\ }\href@noop {} {\bibfield  {journal} {\bibinfo  {journal} {J. Chem.
  Phys.}\ }\textbf {\bibinfo {volume} {90}},\ \bibinfo {pages} {5826} (\bibinfo
  {year} {1989})}\BibitemShut {NoStop}%
\bibitem [{\citenamefont {Celani}, \citenamefont {Puliafito},\ and\
  \citenamefont {Turitsyn}(2005)}]{cela:05}%
  \BibitemOpen
  \bibfield  {author} {\bibinfo {author} {\bibfnamefont {A.}~\bibnamefont
  {Celani}}, \bibinfo {author} {\bibfnamefont {A.}~\bibnamefont {Puliafito}}, \
  and\ \bibinfo {author} {\bibfnamefont {K.}~\bibnamefont {Turitsyn}},\
  }\href@noop {} {\bibfield  {journal} {\bibinfo  {journal} {Europhys. Lett.}\
  }\textbf {\bibinfo {volume} {70}},\ \bibinfo {pages} {464} (\bibinfo {year}
  {2005})}\BibitemShut {NoStop}%
\bibitem [{\citenamefont {Hur}\ and\ \citenamefont {Shaqfeh}(2000)}]{hur:00}%
  \BibitemOpen
  \bibfield  {author} {\bibinfo {author} {\bibfnamefont {J.~S.}\ \bibnamefont
  {Hur}}\ and\ \bibinfo {author} {\bibfnamefont {E.~S.~G.}\ \bibnamefont
  {Shaqfeh}},\ }\href@noop {} {\bibfield  {journal} {\bibinfo  {journal} {J.
  Rheol.}\ }\textbf {\bibinfo {volume} {44}},\ \bibinfo {pages} {713} (\bibinfo
  {year} {2000})}\BibitemShut {NoStop}%
\bibitem [{\citenamefont {Jose}\ and\ \citenamefont {Szamel}(2008)}]{jose:08}%
  \BibitemOpen
  \bibfield  {author} {\bibinfo {author} {\bibfnamefont {P.~P.}\ \bibnamefont
  {Jose}}\ and\ \bibinfo {author} {\bibfnamefont {G.}~\bibnamefont {Szamel}},\
  }\href@noop {} {\bibfield  {journal} {\bibinfo  {journal} {J. Chem. Phys.}\
  }\textbf {\bibinfo {volume} {128}},\ \bibinfo {pages} {224910} (\bibinfo
  {year} {2008})}\BibitemShut {NoStop}%
\bibitem [{\citenamefont {He}\ \emph {et~al.}(2009)\citenamefont {He},
  \citenamefont {Messina}, \citenamefont {L{\"o}wen}, \citenamefont {Kiriy},
  \citenamefont {Bocharova},\ and\ \citenamefont {Stamm}}]{he:09}%
  \BibitemOpen
  \bibfield  {author} {\bibinfo {author} {\bibfnamefont {G.-L.}\ \bibnamefont
  {He}}, \bibinfo {author} {\bibfnamefont {R.}~\bibnamefont {Messina}},
  \bibinfo {author} {\bibfnamefont {H.}~\bibnamefont {L{\"o}wen}}, \bibinfo
  {author} {\bibfnamefont {A.}~\bibnamefont {Kiriy}}, \bibinfo {author}
  {\bibfnamefont {V.}~\bibnamefont {Bocharova}}, \ and\ \bibinfo {author}
  {\bibfnamefont {M.}~\bibnamefont {Stamm}},\ }\href@noop {} {\bibfield
  {journal} {\bibinfo  {journal} {Soft Matter}\ }\textbf {\bibinfo {volume}
  {5}},\ \bibinfo {pages} {3014} (\bibinfo {year} {2009})}\BibitemShut
  {NoStop}%
\bibitem [{\citenamefont {Knudsen}, \citenamefont {de~la Torre},\ and\
  \citenamefont {Elgsaeter}(1996)}]{knud:96}%
  \BibitemOpen
  \bibfield  {author} {\bibinfo {author} {\bibfnamefont {K.~D.}\ \bibnamefont
  {Knudsen}}, \bibinfo {author} {\bibfnamefont {J.~G.}\ \bibnamefont {de~la
  Torre}}, \ and\ \bibinfo {author} {\bibfnamefont {A.}~\bibnamefont
  {Elgsaeter}},\ }\href@noop {} {\bibfield  {journal} {\bibinfo  {journal}
  {Polymer}\ }\textbf {\bibinfo {volume} {37}},\ \bibinfo {pages} {1317}
  (\bibinfo {year} {1996})}\BibitemShut {NoStop}%
\bibitem [{\citenamefont {Lyulin}, \citenamefont {Adolf},\ and\ \citenamefont
  {Davies}(1999)}]{lyul:99}%
  \BibitemOpen
  \bibfield  {author} {\bibinfo {author} {\bibfnamefont {A.~V.}\ \bibnamefont
  {Lyulin}}, \bibinfo {author} {\bibfnamefont {D.~B.}\ \bibnamefont {Adolf}}, \
  and\ \bibinfo {author} {\bibfnamefont {G.~R.}\ \bibnamefont {Davies}},\
  }\href@noop {} {\bibfield  {journal} {\bibinfo  {journal} {J. Chem. Phys.}\
  }\textbf {\bibinfo {volume} {111}},\ \bibinfo {pages} {758} (\bibinfo {year}
  {1999})}\BibitemShut {NoStop}%
\bibitem [{\citenamefont {Jendrejack}, \citenamefont {de~Pablo},\ and\
  \citenamefont {Graham}(2002)}]{jend:02}%
  \BibitemOpen
  \bibfield  {author} {\bibinfo {author} {\bibfnamefont {R.~M.}\ \bibnamefont
  {Jendrejack}}, \bibinfo {author} {\bibfnamefont {J.~J.}\ \bibnamefont
  {de~Pablo}}, \ and\ \bibinfo {author} {\bibfnamefont {M.~D.}\ \bibnamefont
  {Graham}},\ }\href@noop {} {\bibfield  {journal} {\bibinfo  {journal} {J.
  Chem. Phys.}\ }\textbf {\bibinfo {volume} {116}},\ \bibinfo {pages} {7752}
  (\bibinfo {year} {2002})}\BibitemShut {NoStop}%
\bibitem [{\citenamefont {Hsieh}\ and\ \citenamefont {Larson}(2004)}]{hsie:04}%
  \BibitemOpen
  \bibfield  {author} {\bibinfo {author} {\bibfnamefont {C.-C.}\ \bibnamefont
  {Hsieh}}\ and\ \bibinfo {author} {\bibfnamefont {R.~G.}\ \bibnamefont
  {Larson}},\ }\href@noop {} {\bibfield  {journal} {\bibinfo  {journal} {J.
  Rheol.}\ }\textbf {\bibinfo {volume} {48}},\ \bibinfo {pages} {995} (\bibinfo
  {year} {2004})}\BibitemShut {NoStop}%
\bibitem [{\citenamefont {Liu}, \citenamefont {Ashok},\ and\ \citenamefont
  {Muthukumar}(2004)}]{liu:04}%
  \BibitemOpen
  \bibfield  {author} {\bibinfo {author} {\bibfnamefont {S.}~\bibnamefont
  {Liu}}, \bibinfo {author} {\bibfnamefont {B.}~\bibnamefont {Ashok}}, \ and\
  \bibinfo {author} {\bibfnamefont {M.}~\bibnamefont {Muthukumar}},\
  }\href@noop {} {\bibfield  {journal} {\bibinfo  {journal} {Polymer}\ }\textbf
  {\bibinfo {volume} {45}},\ \bibinfo {pages} {1383} (\bibinfo {year}
  {2004})}\BibitemShut {NoStop}%
\bibitem [{\citenamefont {Pamies}\ \emph {et~al.}(2005)\citenamefont {Pamies},
  \citenamefont {Martinez}, \citenamefont {Cifre},\ and\ \citenamefont {de~la
  Torre}}]{pami:05}%
  \BibitemOpen
  \bibfield  {author} {\bibinfo {author} {\bibfnamefont {R.}~\bibnamefont
  {Pamies}}, \bibinfo {author} {\bibfnamefont {M.~C.~L.}\ \bibnamefont
  {Martinez}}, \bibinfo {author} {\bibfnamefont {J.~G.~H.}\ \bibnamefont
  {Cifre}}, \ and\ \bibinfo {author} {\bibfnamefont {J.~G.}\ \bibnamefont
  {de~la Torre}},\ }\href@noop {} {\bibfield  {journal} {\bibinfo  {journal}
  {Macromolecules}\ }\textbf {\bibinfo {volume} {38}},\ \bibinfo {pages} {1371}
  (\bibinfo {year} {2005})}\BibitemShut {NoStop}%
\bibitem [{\citenamefont {Sendner}\ and\ \citenamefont {Netz}(2008)}]{send:08}%
  \BibitemOpen
  \bibfield  {author} {\bibinfo {author} {\bibfnamefont {C.}~\bibnamefont
  {Sendner}}\ and\ \bibinfo {author} {\bibfnamefont {R.~R.}\ \bibnamefont
  {Netz}},\ }\href@noop {} {\bibfield  {journal} {\bibinfo  {journal} {EPL}\
  }\textbf {\bibinfo {volume} {81}},\ \bibinfo {pages} {54006} (\bibinfo {year}
  {2008})}\BibitemShut {NoStop}%
\bibitem [{\citenamefont {Zhang}\ \emph {et~al.}(2009)\citenamefont {Zhang},
  \citenamefont {Donev}, \citenamefont {Weisgraber}, \citenamefont {Alder},
  \citenamefont {Graham},\ and\ \citenamefont {de~Pablo}}]{zhan:09}%
  \BibitemOpen
  \bibfield  {author} {\bibinfo {author} {\bibfnamefont {Y.}~\bibnamefont
  {Zhang}}, \bibinfo {author} {\bibfnamefont {A.}~\bibnamefont {Donev}},
  \bibinfo {author} {\bibfnamefont {T.}~\bibnamefont {Weisgraber}}, \bibinfo
  {author} {\bibfnamefont {B.~J.}\ \bibnamefont {Alder}}, \bibinfo {author}
  {\bibfnamefont {M.~G.}\ \bibnamefont {Graham}}, \ and\ \bibinfo {author}
  {\bibfnamefont {J.~J.}\ \bibnamefont {de~Pablo}},\ }\href@noop {} {\bibfield
  {journal} {\bibinfo  {journal} {J. Chem. Phys.}\ }\textbf {\bibinfo {volume}
  {130}},\ \bibinfo {pages} {234902} (\bibinfo {year} {2009})}\BibitemShut
  {NoStop}%
\bibitem [{\citenamefont {Pierleoni}\ and\ \citenamefont
  {Ryckaert}(1995)}]{pier:95}%
  \BibitemOpen
  \bibfield  {author} {\bibinfo {author} {\bibfnamefont {C.}~\bibnamefont
  {Pierleoni}}\ and\ \bibinfo {author} {\bibfnamefont {J.-P.}\ \bibnamefont
  {Ryckaert}},\ }\href@noop {} {\bibfield  {journal} {\bibinfo  {journal}
  {Macromolecules}\ }\textbf {\bibinfo {volume} {28}},\ \bibinfo {pages} {5097}
  (\bibinfo {year} {1995})}\BibitemShut {NoStop}%
\bibitem [{\citenamefont {Aust}, \citenamefont {Kr{\"o}ger},\ and\
  \citenamefont {Hess}(1999)}]{aust:99}%
  \BibitemOpen
  \bibfield  {author} {\bibinfo {author} {\bibfnamefont {C.}~\bibnamefont
  {Aust}}, \bibinfo {author} {\bibfnamefont {M.}~\bibnamefont {Kr{\"o}ger}}, \
  and\ \bibinfo {author} {\bibfnamefont {S.}~\bibnamefont {Hess}},\ }\href@noop
  {} {\bibfield  {journal} {\bibinfo  {journal} {Macromolecules}\ }\textbf
  {\bibinfo {volume} {32}},\ \bibinfo {pages} {5660} (\bibinfo {year}
  {1999})}\BibitemShut {NoStop}%
\bibitem [{\citenamefont {Gratton}\ and\ \citenamefont
  {Slater}(2005)}]{grat:05}%
  \BibitemOpen
  \bibfield  {author} {\bibinfo {author} {\bibfnamefont {Y.}~\bibnamefont
  {Gratton}}\ and\ \bibinfo {author} {\bibfnamefont {G.~W.}\ \bibnamefont
  {Slater}},\ }\href@noop {} {\bibfield  {journal} {\bibinfo  {journal} {Eur.
  Phys. J. E}\ }\textbf {\bibinfo {volume} {17}},\ \bibinfo {pages} {455}
  (\bibinfo {year} {2005})}\BibitemShut {NoStop}%
\bibitem [{\citenamefont {Ryder}\ and\ \citenamefont
  {Yeomans}(2006)}]{ryde:06}%
  \BibitemOpen
  \bibfield  {author} {\bibinfo {author} {\bibfnamefont {J.~F.}\ \bibnamefont
  {Ryder}}\ and\ \bibinfo {author} {\bibfnamefont {J.~M.}\ \bibnamefont
  {Yeomans}},\ }\href@noop {} {\bibfield  {journal} {\bibinfo  {journal} {J.
  Chem. Phys.}\ }\textbf {\bibinfo {volume} {125}},\ \bibinfo {pages} {194906}
  (\bibinfo {year} {2006})}\BibitemShut {NoStop}%
\bibitem [{\citenamefont {Ripoll}, \citenamefont {Winkler},\ and\ \citenamefont
  {Gompper}(2006)}]{ripo:06}%
  \BibitemOpen
  \bibfield  {author} {\bibinfo {author} {\bibfnamefont {M.}~\bibnamefont
  {Ripoll}}, \bibinfo {author} {\bibfnamefont {R.~G.}\ \bibnamefont {Winkler}},
  \ and\ \bibinfo {author} {\bibfnamefont {G.}~\bibnamefont {Gompper}},\
  }\href@noop {} {\bibfield  {journal} {\bibinfo  {journal} {Phys. Rev. Lett.}\
  }\textbf {\bibinfo {volume} {96}},\ \bibinfo {pages} {188302} (\bibinfo
  {year} {2006})}\BibitemShut {NoStop}%
\bibitem [{\citenamefont {Kobayashi}\ and\ \citenamefont
  {Yamamoto}(2010)}]{koba:10}%
  \BibitemOpen
  \bibfield  {author} {\bibinfo {author} {\bibfnamefont {H.}~\bibnamefont
  {Kobayashi}}\ and\ \bibinfo {author} {\bibfnamefont {R.}~\bibnamefont
  {Yamamoto}},\ }\href@noop {} {\bibfield  {journal} {\bibinfo  {journal}
  {Phys. Rev. E}\ }\textbf {\bibinfo {volume} {81}},\ \bibinfo {pages} {041807}
  (\bibinfo {year} {2010})}\BibitemShut {NoStop}%
\bibitem [{\citenamefont {Huang}\ \emph {et~al.}(2010)\citenamefont {Huang},
  \citenamefont {Winkler}, \citenamefont {Sutmann},\ and\ \citenamefont
  {Gompper}}]{huan:10}%
  \BibitemOpen
  \bibfield  {author} {\bibinfo {author} {\bibfnamefont {C.-C.}\ \bibnamefont
  {Huang}}, \bibinfo {author} {\bibfnamefont {R.~G.}\ \bibnamefont {Winkler}},
  \bibinfo {author} {\bibfnamefont {G.}~\bibnamefont {Sutmann}}, \ and\
  \bibinfo {author} {\bibfnamefont {G.}~\bibnamefont {Gompper}},\ }\href@noop
  {} {\bibfield  {journal} {\bibinfo  {journal} {Macromolecules}\ }\textbf
  {\bibinfo {volume} {43}},\ \bibinfo {pages} {10107} (\bibinfo {year}
  {2010})}\BibitemShut {NoStop}%
\bibitem [{\citenamefont {Huang}\ \emph {et~al.}(2011)\citenamefont {Huang},
  \citenamefont {Sutmann}, \citenamefont {Gompper},\ and\ \citenamefont
  {Winkler}}]{huan:11}%
  \BibitemOpen
  \bibfield  {author} {\bibinfo {author} {\bibfnamefont {C.-C.}\ \bibnamefont
  {Huang}}, \bibinfo {author} {\bibfnamefont {G.}~\bibnamefont {Sutmann}},
  \bibinfo {author} {\bibfnamefont {G.}~\bibnamefont {Gompper}}, \ and\
  \bibinfo {author} {\bibfnamefont {R.~G.}\ \bibnamefont {Winkler}},\
  }\href@noop {} {\bibfield  {journal} {\bibinfo  {journal} {EPL}\ }\textbf
  {\bibinfo {volume} {93}},\ \bibinfo {pages} {54004} (\bibinfo {year}
  {2011})}\BibitemShut {NoStop}%
\bibitem [{\citenamefont {Huang}, \citenamefont {Gompper},\ and\ \citenamefont
  {Winkler}(2012)}]{huan:12}%
  \BibitemOpen
  \bibfield  {author} {\bibinfo {author} {\bibfnamefont {C.-C.}\ \bibnamefont
  {Huang}}, \bibinfo {author} {\bibfnamefont {G.}~\bibnamefont {Gompper}}, \
  and\ \bibinfo {author} {\bibfnamefont {R.~G.}\ \bibnamefont {Winkler}},\
  }\href@noop {} {\bibfield  {journal} {\bibinfo  {journal} {J. Phys.: Condens.
  Matter}\ }\textbf {\bibinfo {volume} {24}},\ \bibinfo {pages} {284131}
  (\bibinfo {year} {2012})}\BibitemShut {NoStop}%
\bibitem [{\citenamefont {Maier}\ and\ \citenamefont
  {R{\"a}dler}(1999)}]{maie:99}%
  \BibitemOpen
  \bibfield  {author} {\bibinfo {author} {\bibfnamefont {B.}~\bibnamefont
  {Maier}}\ and\ \bibinfo {author} {\bibfnamefont {J.~O.}\ \bibnamefont
  {R{\"a}dler}},\ }\href@noop {} {\bibfield  {journal} {\bibinfo  {journal}
  {Phys. Rev. Lett.}\ }\textbf {\bibinfo {volume} {82}},\ \bibinfo {pages}
  {1911} (\bibinfo {year} {1999})}\BibitemShut {NoStop}%
\bibitem [{\citenamefont {Sukhishvili}\ \emph {et~al.}(2000)\citenamefont
  {Sukhishvili}, \citenamefont {M{\"u}ller}, \citenamefont {Gratton},
  \citenamefont {Schweizer},\ and\ \citenamefont {Granick}}]{sukh:00}%
  \BibitemOpen
  \bibfield  {author} {\bibinfo {author} {\bibfnamefont {S.~A.}\ \bibnamefont
  {Sukhishvili}}, \bibinfo {author} {\bibfnamefont {J.~D.}\ \bibnamefont
  {M{\"u}ller}}, \bibinfo {author} {\bibfnamefont {E.}~\bibnamefont {Gratton}},
  \bibinfo {author} {\bibfnamefont {K.~S.}\ \bibnamefont {Schweizer}}, \ and\
  \bibinfo {author} {\bibfnamefont {S.}~\bibnamefont {Granick}},\ }\href@noop
  {} {\bibfield  {journal} {\bibinfo  {journal} {Nature}\ }\textbf {\bibinfo
  {volume} {406}},\ \bibinfo {pages} {146} (\bibinfo {year}
  {2000})}\BibitemShut {NoStop}%
\bibitem [{\citenamefont {Wittmer}\ \emph {et~al.}(2010)\citenamefont
  {Wittmer}, \citenamefont {Meyer}, \citenamefont {Johner}, \citenamefont
  {Kreer},\ and\ \citenamefont {Baschnagel}}]{witt:10}%
  \BibitemOpen
  \bibfield  {author} {\bibinfo {author} {\bibfnamefont {J.~P.}\ \bibnamefont
  {Wittmer}}, \bibinfo {author} {\bibfnamefont {H.}~\bibnamefont {Meyer}},
  \bibinfo {author} {\bibfnamefont {A.}~\bibnamefont {Johner}}, \bibinfo
  {author} {\bibfnamefont {T.}~\bibnamefont {Kreer}}, \ and\ \bibinfo {author}
  {\bibfnamefont {J.}~\bibnamefont {Baschnagel}},\ }\href@noop {} {\bibfield
  {journal} {\bibinfo  {journal} {Phys. Rev. Lett.}\ }\textbf {\bibinfo
  {volume} {105}},\ \bibinfo {pages} {037802} (\bibinfo {year}
  {2010})}\BibitemShut {NoStop}%
\bibitem [{\citenamefont {Chattopadhyay}\ and\ \citenamefont
  {Marenduzzo}(2007)}]{chat:07}%
  \BibitemOpen
  \bibfield  {author} {\bibinfo {author} {\bibfnamefont {A.~K.}\ \bibnamefont
  {Chattopadhyay}}\ and\ \bibinfo {author} {\bibfnamefont {D.}~\bibnamefont
  {Marenduzzo}},\ }\href@noop {} {\bibfield  {journal} {\bibinfo  {journal}
  {Phys. Rev. Lett.}\ }\textbf {\bibinfo {volume} {98}},\ \bibinfo {pages}
  {088101} (\bibinfo {year} {2007})}\BibitemShut {NoStop}%
\bibitem [{\citenamefont {Zhang}, \citenamefont {Li},\ and\ \citenamefont
  {Tang}(2011)}]{zhan:11}%
  \BibitemOpen
  \bibfield  {author} {\bibinfo {author} {\bibfnamefont {Q.}~\bibnamefont
  {Zhang}}, \bibinfo {author} {\bibfnamefont {K.}~\bibnamefont {Li}}, \ and\
  \bibinfo {author} {\bibfnamefont {H.}~\bibnamefont {Tang}},\ }\href@noop {}
  {\bibfield  {journal} {\bibinfo  {journal} {Int. J. Mod. Phys. B}\ }\textbf
  {\bibinfo {volume} {25}},\ \bibinfo {pages} {1899} (\bibinfo {year}
  {2011})}\BibitemShut {NoStop}%
\bibitem [{\citenamefont {Maier}, \citenamefont {Seifert},\ and\ \citenamefont
  {R{\"a}dler}(2002)}]{maie:02}%
  \BibitemOpen
  \bibfield  {author} {\bibinfo {author} {\bibfnamefont {B.}~\bibnamefont
  {Maier}}, \bibinfo {author} {\bibfnamefont {U.}~\bibnamefont {Seifert}}, \
  and\ \bibinfo {author} {\bibfnamefont {J.~O.}\ \bibnamefont {R{\"a}dler}},\
  }\href@noop {} {\bibfield  {journal} {\bibinfo  {journal} {Europhys. Lett.}\
  }\textbf {\bibinfo {volume} {60}},\ \bibinfo {pages} {622} (\bibinfo {year}
  {2002})}\BibitemShut {NoStop}%
\bibitem [{\citenamefont {Ripoll}, \citenamefont {Winkler},\ and\ \citenamefont
  {Gompper}(2007)}]{ripo:07}%
  \BibitemOpen
  \bibfield  {author} {\bibinfo {author} {\bibfnamefont {M.}~\bibnamefont
  {Ripoll}}, \bibinfo {author} {\bibfnamefont {R.~G.}\ \bibnamefont {Winkler}},
  \ and\ \bibinfo {author} {\bibfnamefont {G.}~\bibnamefont {Gompper}},\
  }\href@noop {} {\bibfield  {journal} {\bibinfo  {journal} {Eur. Phys. J. E}\
  }\textbf {\bibinfo {volume} {23}},\ \bibinfo {pages} {349} (\bibinfo {year}
  {2007})}\BibitemShut {NoStop}%
\bibitem [{\citenamefont {Gompper}\ \emph {et~al.}(2009)\citenamefont
  {Gompper}, \citenamefont {Ihle}, \citenamefont {Kroll},\ and\ \citenamefont
  {Winkler}}]{gomp:09}%
  \BibitemOpen
  \bibfield  {author} {\bibinfo {author} {\bibfnamefont {G.}~\bibnamefont
  {Gompper}}, \bibinfo {author} {\bibfnamefont {T.}~\bibnamefont {Ihle}},
  \bibinfo {author} {\bibfnamefont {D.~M.}\ \bibnamefont {Kroll}}, \ and\
  \bibinfo {author} {\bibfnamefont {R.~G.}\ \bibnamefont {Winkler}},\
  }\href@noop {} {\bibfield  {journal} {\bibinfo  {journal} {Adv. Polym. Sci.}\
  }\textbf {\bibinfo {volume} {221}},\ \bibinfo {pages} {1} (\bibinfo {year}
  {2009})}\BibitemShut {NoStop}%
\bibitem [{\citenamefont {Winkler}(2012)}]{wink:12}%
  \BibitemOpen
  \bibfield  {author} {\bibinfo {author} {\bibfnamefont {R.~G.}\ \bibnamefont
  {Winkler}},\ }in\ \href@noop {} {\emph {\bibinfo {booktitle} {Hierarchical
  Methods for Dynamics in Complex Molecular Systems}}},\ Vol.~\bibinfo {volume}
  {10},\ \bibinfo {editor} {edited by\ \bibinfo {editor} {\bibfnamefont
  {J.}~\bibnamefont {Grotenhorst}}, \bibinfo {editor} {\bibfnamefont
  {G.}~\bibnamefont {Sutmann}}, \bibinfo {editor} {\bibfnamefont
  {G.}~\bibnamefont {Gompper}}, \ and\ \bibinfo {editor} {\bibfnamefont
  {D.}~\bibnamefont {Marx}}}\ (\bibinfo  {publisher} {Forschungszentrum
  J{\"u}lich GmbH},\ \bibinfo {address} {J{\"u}lich},\ \bibinfo {year}
  {2012})\BibitemShut {NoStop}%
\bibitem [{\citenamefont {Swope}\ \emph {et~al.}(1982)\citenamefont {Swope},
  \citenamefont {Andersen}, \citenamefont {Berens},\ and\ \citenamefont
  {Wilson}}]{swop:82}%
  \BibitemOpen
  \bibfield  {author} {\bibinfo {author} {\bibfnamefont {W.~C.}\ \bibnamefont
  {Swope}}, \bibinfo {author} {\bibfnamefont {H.~C.}\ \bibnamefont {Andersen}},
  \bibinfo {author} {\bibfnamefont {P.~H.}\ \bibnamefont {Berens}}, \ and\
  \bibinfo {author} {\bibfnamefont {K.~R.}\ \bibnamefont {Wilson}},\
  }\href@noop {} {\bibfield  {journal} {\bibinfo  {journal} {J. Chem. Phys.}\
  }\textbf {\bibinfo {volume} {76}},\ \bibinfo {pages} {637} (\bibinfo {year}
  {1982})}\BibitemShut {NoStop}%
\bibitem [{\citenamefont {Allen}\ and\ \citenamefont
  {Tildesley}(1987)}]{alle:87}%
  \BibitemOpen
  \bibfield  {author} {\bibinfo {author} {\bibfnamefont {M.~P.}\ \bibnamefont
  {Allen}}\ and\ \bibinfo {author} {\bibfnamefont {D.~J.}\ \bibnamefont
  {Tildesley}},\ }\href@noop {} {\emph {\bibinfo {title} {Computer Simulation
  of Liquids}}}\ (\bibinfo  {publisher} {Clarendon Press},\ \bibinfo {address}
  {Oxford},\ \bibinfo {year} {1987})\BibitemShut {NoStop}%
\bibitem [{\citenamefont {Kikuchi}\ \emph {et~al.}(2003)\citenamefont
  {Kikuchi}, \citenamefont {Pooley}, \citenamefont {Ryder},\ and\ \citenamefont
  {Yeomans}}]{kiku:03}%
  \BibitemOpen
  \bibfield  {author} {\bibinfo {author} {\bibfnamefont {N.}~\bibnamefont
  {Kikuchi}}, \bibinfo {author} {\bibfnamefont {C.~M.}\ \bibnamefont {Pooley}},
  \bibinfo {author} {\bibfnamefont {J.~F.}\ \bibnamefont {Ryder}}, \ and\
  \bibinfo {author} {\bibfnamefont {J.~M.}\ \bibnamefont {Yeomans}},\
  }\href@noop {} {\bibfield  {journal} {\bibinfo  {journal} {J. Chem. Phys.}\
  }\textbf {\bibinfo {volume} {119}},\ \bibinfo {pages} {6388} (\bibinfo {year}
  {2003})}\BibitemShut {NoStop}%
\bibitem [{\citenamefont {Ihle}\ and\ \citenamefont {Kroll}(2001)}]{ihle:01}%
  \BibitemOpen
  \bibfield  {author} {\bibinfo {author} {\bibfnamefont {T.}~\bibnamefont
  {Ihle}}\ and\ \bibinfo {author} {\bibfnamefont {D.~M.}\ \bibnamefont
  {Kroll}},\ }\href@noop {} {\bibfield  {journal} {\bibinfo  {journal} {Phys.
  Rev. E}\ }\textbf {\bibinfo {volume} {63}},\ \bibinfo {pages} {020201(R)}
  (\bibinfo {year} {2001})}\BibitemShut {NoStop}%
\bibitem [{\citenamefont {Lamura}\ \emph {et~al.}(2001)\citenamefont {Lamura},
  \citenamefont {Gompper}, \citenamefont {Ihle},\ and\ \citenamefont
  {Kroll}}]{lamu:01}%
  \BibitemOpen
  \bibfield  {author} {\bibinfo {author} {\bibfnamefont {A.}~\bibnamefont
  {Lamura}}, \bibinfo {author} {\bibfnamefont {G.}~\bibnamefont {Gompper}},
  \bibinfo {author} {\bibfnamefont {T.}~\bibnamefont {Ihle}}, \ and\ \bibinfo
  {author} {\bibfnamefont {D.~M.}\ \bibnamefont {Kroll}},\ }\href@noop {}
  {\bibfield  {journal} {\bibinfo  {journal} {Europhys. Lett.}\ }\textbf
  {\bibinfo {volume} {56}},\ \bibinfo {pages} {319} (\bibinfo {year}
  {2001})}\BibitemShut {NoStop}%
\bibitem [{\citenamefont {Lamura}, \citenamefont {Burkhardt},\ and\
  \citenamefont {Gompper}(2001)}]{lamu:01.1}%
  \BibitemOpen
  \bibfield  {author} {\bibinfo {author} {\bibfnamefont {A.}~\bibnamefont
  {Lamura}}, \bibinfo {author} {\bibfnamefont {T.~W.}\ \bibnamefont
  {Burkhardt}}, \ and\ \bibinfo {author} {\bibfnamefont {G.}~\bibnamefont
  {Gompper}},\ }\href@noop {} {\bibfield  {journal} {\bibinfo  {journal} {Phys.
  Rev. E}\ }\textbf {\bibinfo {volume} {64}},\ \bibinfo {pages} {061801}
  (\bibinfo {year} {2001})}\BibitemShut {NoStop}%
\bibitem [{\citenamefont {Hori}, \citenamefont {Prasad},\ and\ \citenamefont
  {Kondev}(2007)}]{hori:07}%
  \BibitemOpen
  \bibfield  {author} {\bibinfo {author} {\bibfnamefont {Y.}~\bibnamefont
  {Hori}}, \bibinfo {author} {\bibfnamefont {A.}~\bibnamefont {Prasad}}, \ and\
  \bibinfo {author} {\bibfnamefont {J.}~\bibnamefont {Kondev}},\ }\href@noop {}
  {\bibfield  {journal} {\bibinfo  {journal} {Phys. Rev. E}\ }\textbf {\bibinfo
  {volume} {75}},\ \bibinfo {pages} {041904} (\bibinfo {year}
  {2007})}\BibitemShut {NoStop}%
\bibitem [{\citenamefont {Kratky}\ and\ \citenamefont {Porod}(1949)}]{krat:49}%
  \BibitemOpen
  \bibfield  {author} {\bibinfo {author} {\bibfnamefont {O.}~\bibnamefont
  {Kratky}}\ and\ \bibinfo {author} {\bibfnamefont {G.}~\bibnamefont {Porod}},\
  }\href@noop {} {\bibfield  {journal} {\bibinfo  {journal} {Recl. Trav. Chim.
  Pays-Bas}\ }\textbf {\bibinfo {volume} {68}},\ \bibinfo {pages} {1106}
  (\bibinfo {year} {1949})}\BibitemShut {NoStop}%
\bibitem [{\citenamefont {Odijk}(1995)}]{odij:95}%
  \BibitemOpen
  \bibfield  {author} {\bibinfo {author} {\bibfnamefont {T.}~\bibnamefont
  {Odijk}},\ }\href@noop {} {\bibfield  {journal} {\bibinfo  {journal}
  {Macromolecules}\ }\textbf {\bibinfo {volume} {28}},\ \bibinfo {pages} {7016}
  (\bibinfo {year} {1995})}\BibitemShut {NoStop}%
\bibitem [{\citenamefont {Livadaru}, \citenamefont {Netz},\ and\ \citenamefont
  {Kreuzer}(2003)}]{liva:03}%
  \BibitemOpen
  \bibfield  {author} {\bibinfo {author} {\bibfnamefont {L.}~\bibnamefont
  {Livadaru}}, \bibinfo {author} {\bibfnamefont {R.~R.}\ \bibnamefont {Netz}},
  \ and\ \bibinfo {author} {\bibfnamefont {H.~J.}\ \bibnamefont {Kreuzer}},\
  }\href@noop {} {\bibfield  {journal} {\bibinfo  {journal} {Macromolecules}\
  }\textbf {\bibinfo {volume} {36}},\ \bibinfo {pages} {3732} (\bibinfo {year}
  {2003})}\BibitemShut {NoStop}%
\bibitem [{\citenamefont {Rosa}\ \emph {et~al.}(2003)\citenamefont {Rosa},
  \citenamefont {Hoang}, \citenamefont {Marenduzzo},\ and\ \citenamefont
  {Maritan}}]{rosa:03}%
  \BibitemOpen
  \bibfield  {author} {\bibinfo {author} {\bibfnamefont {A.}~\bibnamefont
  {Rosa}}, \bibinfo {author} {\bibfnamefont {T.~X.}\ \bibnamefont {Hoang}},
  \bibinfo {author} {\bibfnamefont {D.}~\bibnamefont {Marenduzzo}}, \ and\
  \bibinfo {author} {\bibfnamefont {A.}~\bibnamefont {Maritan}},\ }\href@noop
  {} {\bibfield  {journal} {\bibinfo  {journal} {Macromolecules}\ }\textbf
  {\bibinfo {volume} {36}},\ \bibinfo {pages} {10095} (\bibinfo {year}
  {2003})}\BibitemShut {NoStop}%
\bibitem [{sup()}]{supp}%
  \BibitemOpen
  \href@noop {} {}\bibinfo {howpublished} {See supplementary material at [URL
  will be inserted by AIP] for movies of a semiflexible polymer ($L_p/L=10$) at
  $Wi=8$ and $Wi=800$.}\BibitemShut {Stop}%
\bibitem [{\citenamefont {Lattanzi}, \citenamefont {Munk},\ and\ \citenamefont
  {Frey}(2004)}]{latt:04}%
  \BibitemOpen
  \bibfield  {author} {\bibinfo {author} {\bibfnamefont {G.}~\bibnamefont
  {Lattanzi}}, \bibinfo {author} {\bibfnamefont {T.}~\bibnamefont {Munk}}, \
  and\ \bibinfo {author} {\bibfnamefont {E.}~\bibnamefont {Frey}},\ }\href@noop
  {} {\bibfield  {journal} {\bibinfo  {journal} {Phys. Rev. E}\ }\textbf
  {\bibinfo {volume} {69}},\ \bibinfo {pages} {021801} (\bibinfo {year}
  {2004})}\BibitemShut {NoStop}%
\bibitem [{\citenamefont {Steinhauser}, \citenamefont {K{\"o}ster},\ and\
  \citenamefont {Pfohl}(2012)}]{stei:12}%
  \BibitemOpen
  \bibfield  {author} {\bibinfo {author} {\bibfnamefont {D.}~\bibnamefont
  {Steinhauser}}, \bibinfo {author} {\bibfnamefont {S.}~\bibnamefont
  {K{\"o}ster}}, \ and\ \bibinfo {author} {\bibfnamefont {T.}~\bibnamefont
  {Pfohl}},\ }\href@noop {} {\bibfield  {journal} {\bibinfo  {journal} {ACS
  Macro Letters}\ }\textbf {\bibinfo {volume} {1}},\ \bibinfo {pages} {541}
  (\bibinfo {year} {2012})}\BibitemShut {NoStop}%
\bibitem [{\citenamefont {Chelakkot}, \citenamefont {Winkler},\ and\
  \citenamefont {Gompper}(2010)}]{chel:10}%
  \BibitemOpen
  \bibfield  {author} {\bibinfo {author} {\bibfnamefont {R.}~\bibnamefont
  {Chelakkot}}, \bibinfo {author} {\bibfnamefont {R.~G.}\ \bibnamefont
  {Winkler}}, \ and\ \bibinfo {author} {\bibfnamefont {G.}~\bibnamefont
  {Gompper}},\ }\href@noop {} {\bibfield  {journal} {\bibinfo  {journal} {EPL}\
  }\textbf {\bibinfo {volume} {91}},\ \bibinfo {pages} {14001} (\bibinfo {year}
  {2010})}\BibitemShut {NoStop}%
\bibitem [{\citenamefont {Chelakkot}, \citenamefont {Winkler},\ and\
  \citenamefont {Gompper}(2011)}]{chel:11}%
  \BibitemOpen
  \bibfield  {author} {\bibinfo {author} {\bibfnamefont {R.}~\bibnamefont
  {Chelakkot}}, \bibinfo {author} {\bibfnamefont {R.~G.}\ \bibnamefont
  {Winkler}}, \ and\ \bibinfo {author} {\bibfnamefont {G.}~\bibnamefont
  {Gompper}},\ }\href@noop {} {\bibfield  {journal} {\bibinfo  {journal} {J.
  Phys.: Condens. Matter}\ }\textbf {\bibinfo {volume} {23}},\ \bibinfo {pages}
  {184117} (\bibinfo {year} {2011})}\BibitemShut {NoStop}%
\bibitem [{\citenamefont {Delgado-Buscalioni}(2006)}]{delg:06}%
  \BibitemOpen
  \bibfield  {author} {\bibinfo {author} {\bibfnamefont {R.}~\bibnamefont
  {Delgado-Buscalioni}},\ }\href@noop {} {\bibfield  {journal} {\bibinfo
  {journal} {Phys. Rev. Lett.}\ }\textbf {\bibinfo {volume} {96}},\ \bibinfo
  {pages} {088303} (\bibinfo {year} {2006})}\BibitemShut {NoStop}%
\end{thebibliography}

%

\end{document}